\documentclass[aps,10pt,twocolumn,tightenlines]{revtex4-1}
\usepackage{graphicx}
\usepackage[space]{grffile}
\usepackage{latexsym}
\usepackage{amsfonts,amsmath,amssymb}
\usepackage{url}
\usepackage[utf8]{inputenc}
\usepackage{hyperref}
\hypersetup{colorlinks=false,pdfborder={0 0 0}}
\usepackage{textcomp}
\usepackage{longtable}
\usepackage{multirow,booktabs}
\usepackage[utf8]{inputenc} 
\newcommand{\Qbar}{\overline{Q}}

\usepackage{tikz}
\usetikzlibrary{%
    decorations.pathreplacing,%
    decorations.pathmorphing%
}
\usepackage{mathtools}
\usetikzlibrary{calc}

\newcommand{\Cross}{$\mathbin{\tikz [x=1.4ex,y=1.4ex,line width=2, black] \draw (0,0) -- (1,1) (0,1) -- (1,0);}$}%
\newcommand{\Crossx}{$\mathbin{\tikz [x=1.4ex,y=1.4ex, line width=2, black] \draw (0,2) -- (2,2) (1.,3.) -- (1.,1);}$}%

\begin{document}

\title{Tau energy loss and ultrahigh energy skimming tau neutrinos}

\author{Yu Seon Jeong}
\email{ysjeong@email.arizona.edu}
\affiliation{National Institute of Supercomputing and Networking, KISTI, Daejeon 34141, Korea \\
 Department of Physics, University of Arizona, Tucson, AZ 85721}
\author{Minh Vu Luu}
\email{minh-luu@uiowa.edu}
\affiliation{Department of Physics and Astronomy, University of Iowa, Iowa City, IA 52242}
\author{Mary Hall Reno}
\email{mary-hall-reno@uiowa.edu}
\affiliation{Department of Physics and Astronomy, University of Iowa, Iowa City, IA 52242}
\author{Ina Sarcevic}
\email{ina@physics.arizona.edu}
\affiliation{Department of Physics, University of Arizona, Tucson, AZ 85721\\ 
Department of Astronomy, University of Arizona, Tucson, AZ 85721}
\date{\today}

\begin{abstract}
We consider propagation of high energy earth-skimming taus 
produced in interactions of astrophysical tau neutrinos.  
For astrophysical tau neutrinos we take generic  power-law flux, 
$E^{-2}$ and the cosmogenic flux initiated by the protons.  
We calculate tau energy loss in several approaches, such as 
dipole models and the phenomenological approach in which 
parameterization of the $F_2$ is used.  
We evaluate the tau 
neutrino charged-current cross 
section using the same approaches for consistency.  We find 
that uncertainty in the neutrino cross section and in 
the tau energy loss partially compensate giving very small 
theoretical uncertainty in  the emerging tau flux for 
distances ranging from $2$ km to $100$ km and 
for the energy range between $10^6$ GeV and $10^{11}$ GeV, focusing on energies above $10^8$ GeV.  
When we consider uncertainties in 
the neutrino cross section, inelasticity in neutrino interactions 
and the tau energy loss, 
which are not correlated, i.e. they are not 
all calculated in the same approach, theoretical 
uncertainty ranges from 
about $30\%$ and  $60 \%$ at $10^8$ GeV to about factors of 3.3 and 3.8 
at $10^{11}$ GeV for the $E^{-2}$ flux and the 
 cosmogenic flux, respectively, for the distance of 10 km rock. The spread in predictions significantly increases for
 much larger distances, e.g., $\sim 1,000$ km. Most of the uncertainty comes from the treatment of photonuclear interactions of the tau
 in transit through large distances.
We also consider Monte Carlo calculation of the tau 
propagation and we find that the result for the emerging 
tau flux is in agreement with the result obtained using 
analytic approach.  Our results are relevant to 
several experiments that are looking for skimming 
astrophysical taus, such as the Pierre Auger Observatory, HAWC and Ashra.  
We evaluate the aperture for the Auger and discuss 
briefly application to the the other two experiments.  
\end{abstract} 

\maketitle

\section{Introduction}

Astrophysical sources that produce observed 
high energy cosmic rays and photons should also be sources of neutrinos (see, e.g., Refs. \cite{Gaisser:1994yf,Katz:2011ke,Anchordoqui:2013dnh}). While individual neutrino
sources have not yet been detected, the IceCube Neutrino Observatory 
has evidence of a diffuse flux of neutrinos that lies above the background
of neutrinos produced in the atmosphere for neutrino 
energies above $\sim 10^5-10^6$ GeV \cite{Aartsen:2013eka}. Further characterization of the diffuse flux is underway at IceCube and at underwater
observatories \cite{Fusco:2016vil}.  

Detection of neutrinos from astrophysical sources is expected 
to reveal details about the site of cosmic ray acceleration and about the 
propagation of cosmic rays through the thermal cosmic background. 
The feature that neutrinos can escape sources without interacting and are not 
 affected by magnetic fields translates to a challenge in neutrino detection.  
The underground observatories in ice and water instrument a large volume of target material that is transparent to the light signals that neutrino weak interactions produce. At higher energies, the neutrino interaction probabilities increase, however, the expected flux of neutrinos decreases more quickly as a function of energy, requiring even larger target volumes.

A well-known technique to overcome the problem of large target volumes for 
ultrahigh energy neutrinos has been explored for detecting 
astrophysical tau neutrinos 
 in air shower experiments. 
The idea is to use the Earth as a neutrino converter then to use an 
emerging tau decay into a shower to detect the ultrahigh energy tau neutrino flux \cite{Domokos:1997ve,Fargion_2004,Fargion:2000iz,Fargion:2003kn,Athar:2000rx,Feng:2001ue,Kusenko:2001gj,Hou:2002bh,Tseng:2003pn,Aramo:2004pr,PalomaresRuiz:2005xw}.
Currently, the Pierre Auger Observatory has set limits on the 
tau neutrino flux in the energy range $2\times 10^{8}$ GeV up to 
$2\times 10^{10}$ GeV 
\cite{Abraham:2009uy, Abreu:2012zz} using this technique. 
The potential to use the volcano adjacent to HAWC as the neutrino 
converter has also been explored \cite{Vargas:2016hcp}. 
In addition, there is a proposal for the Ashra detector that would also use a 
mountain as a neutrino converter, 
for example Mauna Kea \cite{Asaoka:2012em}.

Among the uncertainties in these tau neutrino flux limits or 
eventual  observation are three elements: 
the uncertainty in the high energy neutrino cross section \cite{Gandhi:1998ri,Reno:2004cx,Jeong:2010za,Goncalves:2010ay,CooperSarkar:2011pa,Fiore:2011gx,Goncalves:2015fua,Albacete:2015zra,Arguelles:2015wba}
required to convert the tau neutrino
to the tau lepton in the Earth, the inelasticity for 
neutrino scattering with isoscalar nucleons (how much of the neutrino energy is transferred to the tau) and the 
tau electromagnetic loss \cite{Dutta:2000hh,Chirkin:2004hz,Dutta:2005yt,Armesto:2007tg,Bigas:2008ff,Koehne:2013gpa} as the
tau transits the remaining rock before it emerges to decay into an air shower. 

Tau energy loss is through electromagnetic interactions via ionization, bremsstrahlung, pair production and photonuclear interactions, e.g., see Refs. \cite{Dutta:2000hh,Chirkin:2004hz,Dutta:2005yt,Armesto:2007tg,Lipari:1991ut,Lohmann:1985qg,Antonioli:1997qw,Koehne:2013gpa,Bigas:2008ff,Sokalski:2000nb}.
At high energies, the ionization energy loss is negligible. For taus, the photonuclear  and pair production processes are much larger than bremsstrahlung, and
as the energy increases, the photonuclear interaction dominates. For muons, bremsstrahlung, pair production and photonuclear interactions give comparable contributions to the electromagnetic energy loss \cite{Lohmann:1985qg}. The uncertainties in the high energy behavior of the photonuclear interaction depend 
on the  parameterizations and assumptions about the nucleon structure. 
The neutrino cross section uncertainties are also related to the 
nucleon structure at higher momentum transfers. 

In this paper, we focus on two goals. The first is to evaluate the 
theoretical uncertainties in the photonuclear energy loss, neutrino cross sections and neutrino inelasticity in a way that correlates the assumptions about the nuclear structure for both scattering processes. We consider energies between $10^6-10^{12}$ GeV, for taus and their corresponding neutrinos that come from cosmic sources 
and focus more on energies above $10^8$ GeV. 
We use an input spectrum of neutrinos scaling with energy as $E^{-2}$ as a representative neutrino flux.  In addition, we consider a cosmogenic neutrino 
flux which is the flux from the interactions of cosmic rays with the 
cosmic neutrino background (GZK neutrinos).  Cosmic rays have been detected up to $ 3 \times 10^{20}$ eV$= 3 \times 10^{11}$ GeV \cite{Bird:1994wp}, 
and the shape of the cosmogenic neutrino flux is deduced from those 
observations.

Our second goal is to provide a semi-analytic approximation to the stochastic evaluation of first, the neutrino interaction, then the tau energy loss in matter. Uncertainties from a continuous versus stochastic treatment of the tau energy loss
are discussed here (see also \cite{Bigas:2008ff}).
We examine the impact of stochastic versus
approximate treatments of the neutrino interaction itself as well. The semi-analytic treatment reliably allows a survey of the impact on the tau flux of multiple approaches to calculating the neutrino (differential) cross section and tau energy loss.

We begin the paper with a summary of electromagnetic energy loss 
of charged leptons, focusing on the photonuclear process and 
the range of predictions that come from different approaches 
to modeling the electromagnetic structure function $F_2$ beyond 
the range of experimental measurements.  To 
exhibit the effectiveness of analytic approximations to the tau energy 
loss we show an example for an incident tau lepton flux given by 
power-law, $E^{-2}$.  
Section 3 connects $F_2$ from electromagnetic scattering of charged leptons to the high energy neutrino cross section. In Section 4, we show results for incident neutrino fluxes characterized by a power law and for a representative cosmogenic neutrino flux.  We discuss applications to the Pierre Auger Observatory in Section 5. Our conclusions are in Section 6.

\section{Electromagnetic energy loss of charged leptons}

\subsection{Average energy loss}

The propagation of charged leptons through materials can be described through the average energy loss through electromagnetic processes. The
average energy loss is usually written as
\begin{equation}
\Biggl\langle \frac{dE}{dX}\Biggr\rangle = -(\alpha+\beta E)\ ,
\end{equation}
in terms of the column depth
\begin{equation}
X(D) = \int_0^D \rho(\ell) d\ell 
\end{equation}
for density $\rho$ and distance $D$.
The parameter $\alpha$ describes ionization energy loss from the Bethe-Block formula \cite{Rossi:1952kt}, weakly dependent on energy, and at high energies, weakly dependent on lepton mass $m_\ell$. For $E=10^6-10^9$ GeV,
$\alpha = 3.1-3.6\times 10^{-3}$ GeV\,cm$^2$/g. At high energies, above  a muon energy $E\sim 10^3 $ GeV and 
tau energy $E\sim 10^4$ GeV, one can ignore the energy independent ionization energy loss, which we will do
here. 

The three contributions from electromagnetic energy loss to $\beta$ come from bremsstrahlung, electron-positron pair production
and the photonuclear (inelastic scattering) processes. They are written as
\begin{equation}
\beta = \beta^{\rm brem}+\beta^{\rm pair} + \beta^{\rm nuc}\ ,
\end{equation}
where $\beta$ is calculated for each process by
\begin{equation}
\label{eq:betadef}
\beta(E) = \frac{N}{A}\int dy\, y\, \frac{d\sigma(y,E)}{dy}
\end{equation}
in terms of the inelasticity parameter $y = (E-E')/E$ for incident energy $E$ and outgoing energy $E'$.
As discussed below, $\beta(E)$ depends on the charged lepton, so we label it with $\beta_\ell$ for $\ell=\mu$ or
$\ell=\tau$.

We evaluate the bremsstrahlung energy loss following Petrukhin and Sestakov \cite{pesh:1968}. 
Pair production energy loss follows Ref. \cite{kp}. As shown in Ref. \cite{Koehne:2013gpa} (see also Ref. \cite{Klein:1998du}), suppression of pair production and bremsstrahlung by the Landau-Pomeranchuk-Migdal and Ter-Mikaelian effects from coherence effects and dielectric suppression do not affect muon (and tau) energy loss below $E_\ell =10^{12}$ GeV, so we do not discuss it here. 
Scaling of the energy loss dependence on the charged lepton mass is
discussed in Refs. \cite{Tannenbaum:1990ae,Reno:2005si}. For bremsstrahlung, $\beta_\ell \sim 1/m_\ell^2$, while
for pair production, $\beta_\ell\sim 1/m_\ell$.

The photonuclear process also scales as $\beta_\ell \sim 1/m_\ell$. The photon exchange between charged lepton and nucleon probes the nucleon structure.
The differential cross section in 
Eq. (\ref{eq:betadef}) depends on the nucleon structure functions $F_1(x,Q^2)$ and $F_2(x,Q^2)$ which depend on the 
lepton four momentum transfer squared $Q^2=-(\ell - \ell ')^2=-q^2$ and $x=Q^2/(2 p\cdot q)$ for
\begin{equation}
\ell + N(p) \to \ell' +X,\quad q=\ell - \ell'\ .
\end{equation}
The differential distribution for the photonuclear interaction is
\begin{eqnarray} 
\nonumber
\frac{d^2\sigma^{\rm nuc}}{dx\, dQ^2} &=& \frac{4\pi \alpha^2}{Q^4}
\Biggl[\Biggl(1-y-\frac{Mxy}{2E}\Biggr) \frac{F_2(x,Q^2)}{x}\\
\nonumber
&+&\Biggl( 1-\frac{2m^2}{Q^2}\Biggr) y^2 F_1(x,Q^2)
\Biggr]\\
\nonumber
&=& \frac{4\pi \alpha^2}{Q^4}\frac{F_2(x,Q^2)}{x}
\Biggl[ 1-y-\frac{Mxy}{2E}\\
\nonumber
&+& \Biggl( 1-\frac{2m^2}{Q^2}\Biggr) \frac{y^2}{2}\\
&\times & \Biggl(
1+\frac{4M^2x^2}{Q^2}-\frac{F_L(x,Q^2)}{F_2(x,Q^2)}\Biggr)
\Biggr] \ ,\\
F_L &=& \Biggl(1+\frac{4 M^2 x^2}{Q^2}\Biggr) F_2 - 2 x F_1
\end{eqnarray}
The full $x$ and $Q^2$ dependence of $F_2(x,Q^2)$ in the photonuclear electromagnetic interaction gives an energy dependence to $\beta_\ell^{\rm nuc}(E)$ \cite{Dutta:2000hh}. 

The dependence on the structure functions, in particular at small values of $x$ and  low $Q^2$, introduces an uncertainty in $\beta_\ell^{\rm nuc}$.
One option is to use a direct parameterization of $F_2(x,Q^2)$, and to use the leading order Callan-Gross relation 
$2xF_1=F_2$ to obtain $F_1(x,Q^2)$. Corrections can be implemented with $R=F_L/(2xF_1)$. The dependence of
the photonuclear energy loss on a reasonable range of $R$ was shown to be small in Ref. \cite{Koehne:2013gpa}.  In the cases of direct parameterizations, we set $R=0$. Using structure functions based on parton model
evaluations is not possible here, even with small-$x$ extrapolations, because of the low $Q^2$ required.

Photonuclear interactions, in general,  
 consist of the soft component (non-perturbative)
and the hard component (perturbative).
Early evaluations (see, e.g., Ref. \cite{ Lohmann:1985qg}) of $\beta_\ell^{\rm nuc}$ relied on a $Q$ independent parameterization by Bezrukov and Bugaev in Ref. \cite{Bezrukov:1981ci}. 
Bezrukov and Bugaev modeled the soft component of the structure functions with a modified generalized
vector dominance model.
In Ref. \cite{Dutta:2000hh}, the impact of the full range of $Q^2$ in the evaluation of $\beta_\ell^{\rm nuc}$ was discussed using
the structure functions parameterization of Abramowicz et al. (ALLM) \cite{Abramowicz:1991xz,Abramowicz:1997ms}. A similar parameterization of
Butkevich and Mikhailov \cite{Butkevich:2001aw}
based on the Capella et al. model (CKMT) \cite{Capella:1994cr} has been used in Ref. \cite{Chirkin:2004hz,Koehne:2013gpa}. 
The Bezrukov and Bugaev result was augmented by mass corrections and a hard component that depends on $Q^2$  in Ref.
\cite{Bugaev:2002gy,*Bugaev:2003sw}, giving similar results to the ALLM 
parameterization. More recently, a new parameterization of $F_2$ by Block et al. (BDHM) \cite{Block:2014kza}, guided by unitarity
considerations \cite{Block:2013mia,Block:2013nia}, has been provided. It has a somewhat different behavior at low-$x$ than the
ALLM parameterization. We show below that the difference in 
small-$x$ behavior of
the BDHM and ALLM parameterizations of $F_2$ have implications for the high energy flux of taus emerging from the Earth. 
 
A QCD-motivated
approach to the photonuclear electromagnetic interactions has the photon, given its wave function,
 split into a  $q\bar{q}$-pair which then interacts with the nucleon non-perturbatively for the
case when their transverse separation is large and perturbatively when
it is small.   
This so-called dipole model \cite{Nikolaev:1990ja,Mueller:1993rr}, while based on perturbation theory, can be used as
an approximate calculation even at low $Q^2$.
The structure function therefore depends on the photon wave function (squared) which itself depends on the transverse separation $r$ and longitudinal
momentum fraction $z$, and the dipole cross section $\sigma_{\rm dip}$, implicitly summed over quark flavors:
\begin{eqnarray}
\nonumber
F_2(x,Q^2) &=& \frac{Q^2}{4\pi^2 \alpha} \int d^2 r\int dz\biggl[ |\psi_L^\gamma(z,r)|^2 \\
&+&
|\psi_T^\gamma(z,r)|^2\biggr] \sigma_{\rm dip}(x,r)\ .
\label{eq:F2dipole}
\end{eqnarray}
Formulae for the wave functions squared are listed in the appendix. In the dipole model, the longitudinal structure function can be included explicitly as well. Different models are used to provide the dipole
cross section. We use the dipole cross section of
Soyez \cite{Soyez:2007kg}. The Soyez dipole cross section
is based on a functional form discussed by Iancu, Itakura and Munier \cite{Iancu:2003ge} from an approximate
solution to the non-linear Balitsky-Kovchegov (BK) equations \cite{Balitsky:1995ub,Kovchegov:1999yj}. Dipole cross section
parameters were fit to $F_2$ data using 
Eq. (\ref{eq:F2dipole}) for $0.045$ GeV$^2\leq Q^2\leq 150$ GeV$^2$ and $10^{-6}\leq x\leq 10^{-2}$.
Similar results are obtained with the dipole of Albacete et al. (AAMQS) \cite{Albacete:2015zra}. The
AAMQS dipole includes further theoretical improvements to the approximate solution to the BK equations including the running coupling constant.

\begin{figure}[htb]
\centering
	\includegraphics[width=\columnwidth]{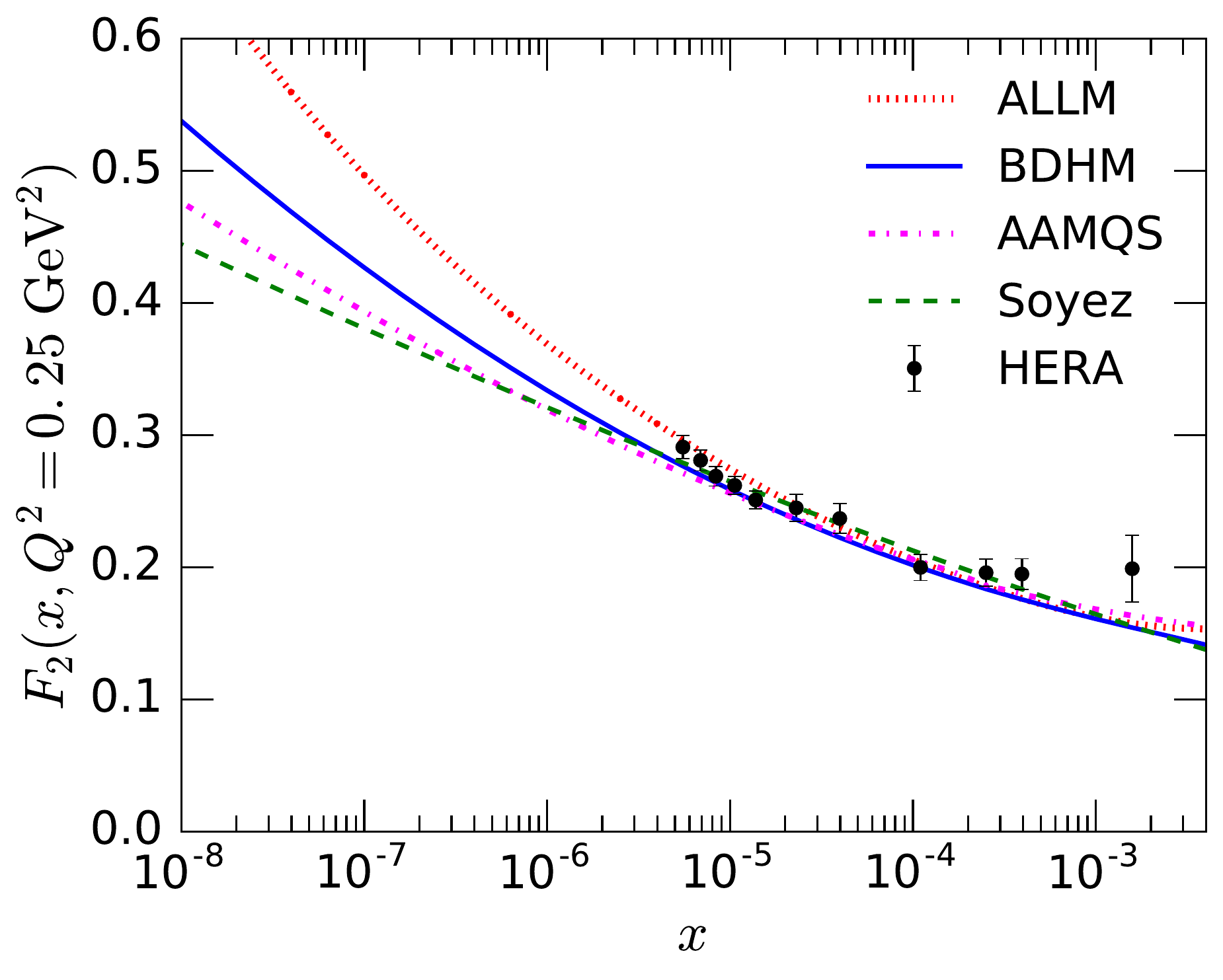}
	\includegraphics[width=\columnwidth]{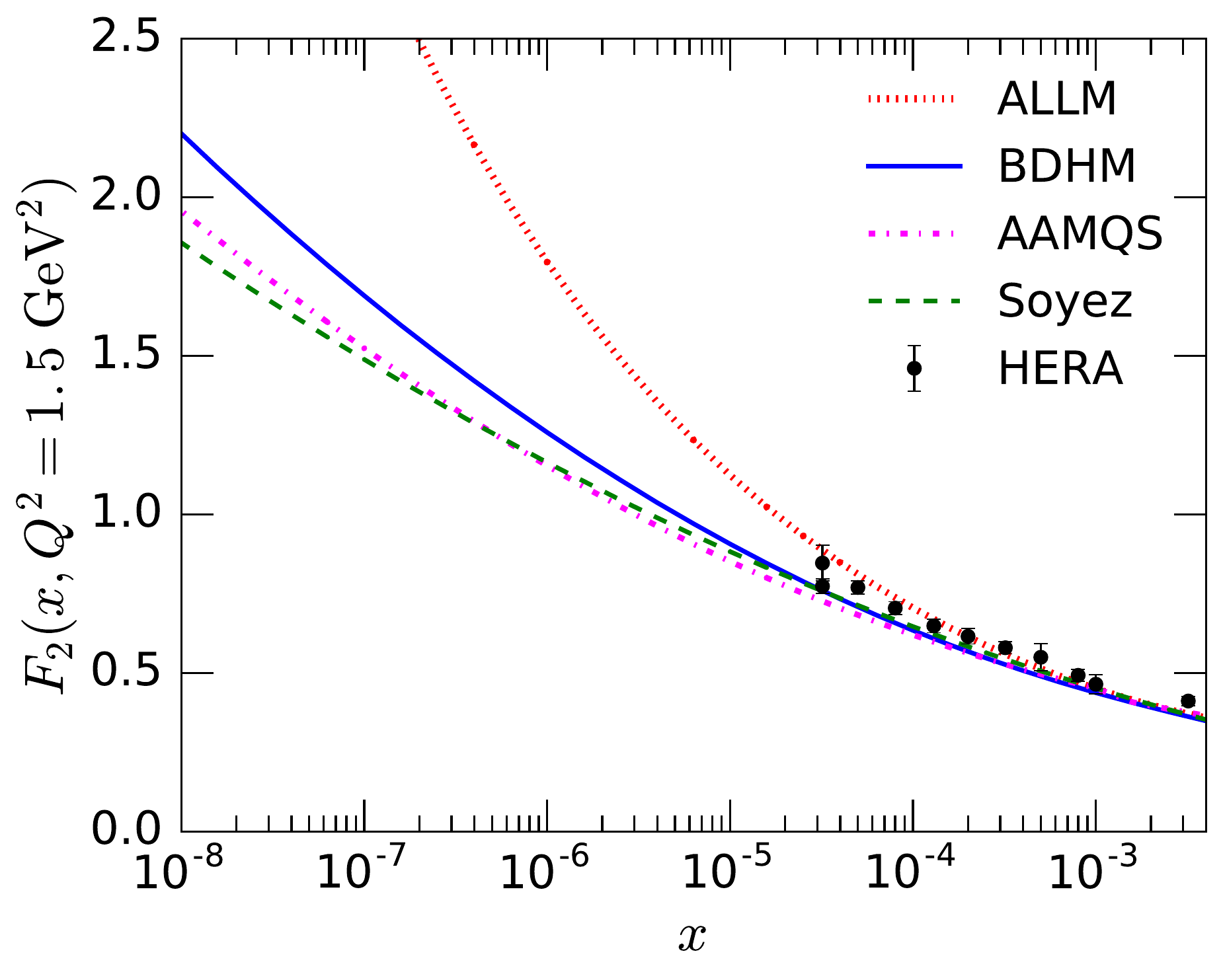}
	
	\caption{The electromagnetic structure function $F_2(x,Q^2)$ for $Q^2=0.25$ GeV$^2$ (upper) and
	1.5 GeV$^2$ (lower) evaluated using 
	the ALLM \cite{Abramowicz:1997ms}, 
	and BDHM \cite{Block:2014kza} parameterizations, and two dipole models (Soyez \cite{Soyez:2007kg} and
	AAMQS \cite{Albacete:2015zra}). The data are combined HERA data from H1 and ZEUS \cite{Aaron:2009aa}.}
	\label{fig:F2}
\end{figure}

Each of the approaches gives a prediction for the small-$x$ electromagnetic structure function for $x$ below the measured regime. Fig. \ref{fig:F2} shows four approaches for $Q^2=0.25$ GeV$^2$ and $Q^2=1.5$ GeV$^2$ , along with HERA data combined from H1 and ZEUS \cite{Aaron:2009aa}.   The ALLM parameterization of $F_2$ has a steeper growth with 
small-$x$ than the other approaches. The BDHM parameterization, the Soyez dipole and AAMQS dipole do well in comparison with the small-$x$ HERA data and have similar small-$x$ behaviors. 
Since the Soyez and AAMQS dipoles have similar small-$x$ behavior, we use the Soyez dipole as representative of dipole models in our evaluation of the tau fluxes below. 

It may be eventually possible to distinguish between the ALLM parameterizations and other approaches from LHC data for low invariant mass
and forward rapidity kinematic regions. 
These data constrain low-$x$ parton distribution functions, as discussed in, e.g., Ref.
\cite{Gauld:2016kpd}. At least for sufficiently large $Q^2$, the parton distribution functions can be combined to
form the electromagnetic structure functions at smaller $x$ values than accessible with HERA, where the ALLM
curve is distinct from the other curves in Fig. \ref{fig:F2}.

For $\beta_\ell^{\rm nuc}$ for rock, with $Z=11$ and $A=22$, nuclear corrections are required. 
One can define
\begin{equation}
S (A,x,Q^2) = \frac{F_2^A(x,Q^2)}{A F_2(x, Q^2)}\ ,
\end{equation}
the nuclear structure function normalized by $A$ times the proton structure function.
For the ALLM and BDHM parameterizations, we use a shadow factor  of the form $S(A,x,Q^2)\simeq
S(A,x)$,
\[
S(A,x)
= \left\{ \begin{array} {cl}
A^{-0.1} & x<0.0014 \\
A^{0.069\log_{10}x+0.097 }& 0.0014<x<0.04 \\
1 & 0.04<x
\end{array}\right.
\]
based on Fermilab E665 data \cite{Adams:1992vm,Adams:1992nf,Adams:1995is}. For the dipole model, one can use the
Glauber-Gribov (GG) prescription to incorporate the dipole cross section for 
 protons to the one for nuclei, where 
\begin{eqnarray}
\sigma_{\rm dip}^A(x,r)&=&\int d^2\vec{b}\, \sigma_{\rm dip}^A(x,r,b)\, ,\\
\sigma_{\rm dip}^A(x,r,b) &=& 2\Biggl[ 1-\exp\Biggl( -\frac{1}{2} AT_A(b)\sigma_{\rm dip}(x,r)\Biggr)\Biggr] .
\end{eqnarray}
This requires the nuclear density $\rho_A$ and nuclear profile $T_A(b)$, normalized to unity, with
\begin{eqnarray}
\rho_A (z,\vec{b}) &= &\frac{1}{\pi^{3/2} a^3} e^{-r^2/a^2},\   r^2=z^2+\vec{b}^{\, 2}\\
T_A(b) &=& \int dz\rho_A(z,\vec{b})\, ,\\
\int d^2 \vec{b}\, T_A(b)&=& 1\ ,
\end{eqnarray}
and $a^2=2 R_A^2/3$. The quantity $R_A$ is the nuclear radius $R_A = 1.12\, A^{1/3}-0.86/A^{1/3}$ fm.
Finally, for the Soyez dipole, nuclear effects can be included with a modification of the saturation scale as discussed 
by 
Armesto, Salgado and Wiedemann (ASW) \cite{Armesto:2004ud}.

\begin{figure}[htb]
\centering
	\includegraphics[width=\columnwidth]{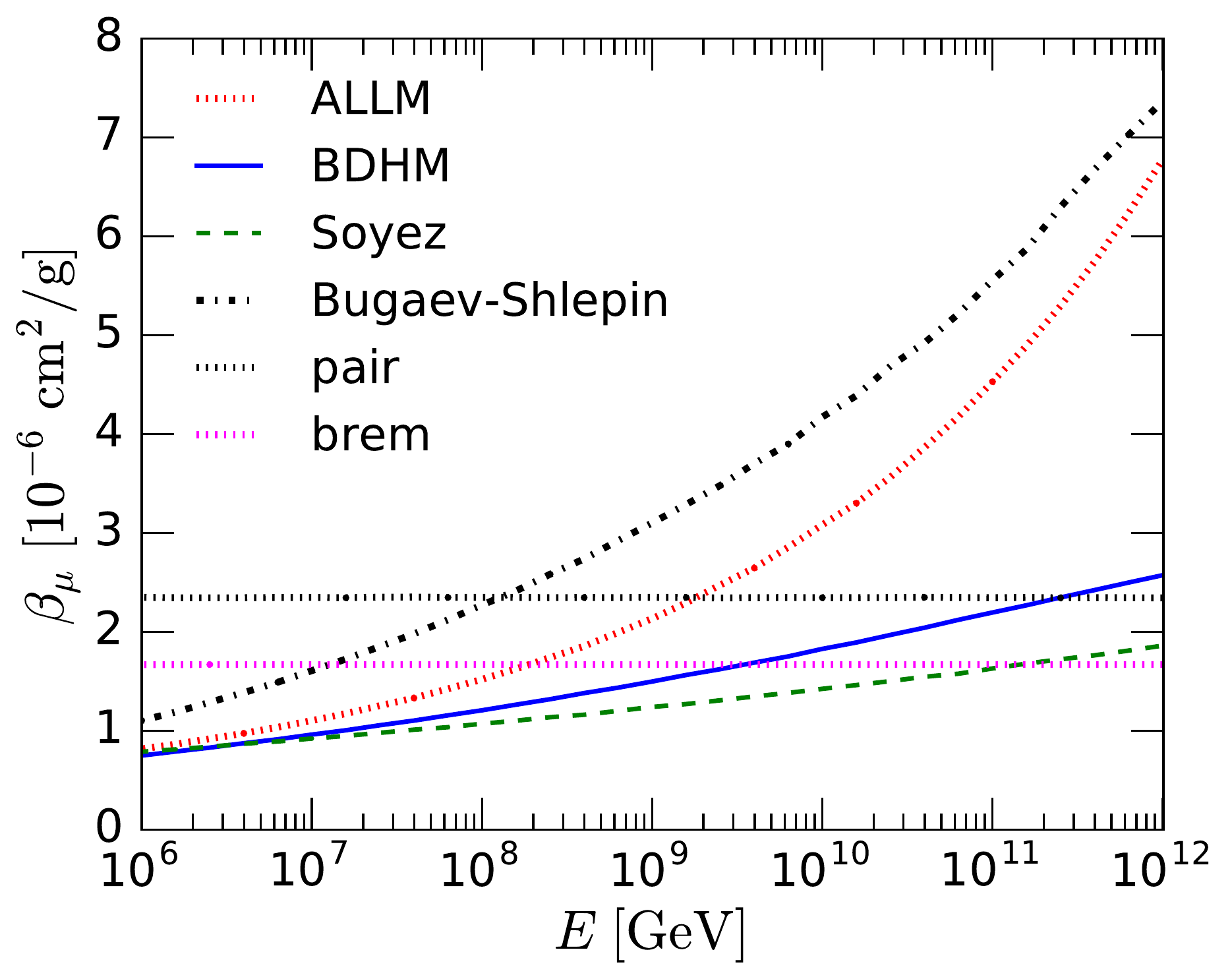}
	\caption{ For muons, $\beta_\mu$ vs the energy of the muon $E$ for rock. The $\beta_\mu^{\rm nuc}$ for the dipole model (Soyez) is shown using dashed lines,  and parameterizations of $F_2$  by ALLM (dotted), and BDHM (solid). 
The dot-dashed line comes from Bugaev and Shlepin \cite{Bugaev:2002gy}. The pair production $\beta_\mu^{\rm pair}$ and bremsstrahlung $\beta_\mu^{\rm brem}$ are also shown with the upper and lower dotted lines, respectively. For bremsstrahlung, $\beta_\mu^{\rm brem} =1.67\times 10^{-6}$ cm$^2$/g and for pair production,
$\beta_\mu^{\rm pair}=2.35\times 10^{-6}$  cm$^2$/g, essentially independent of energy for this energy range.}
	\label{fig:betamuon}
\end{figure}

\begin{figure}[htb]
\centering
	\includegraphics[width=\columnwidth]{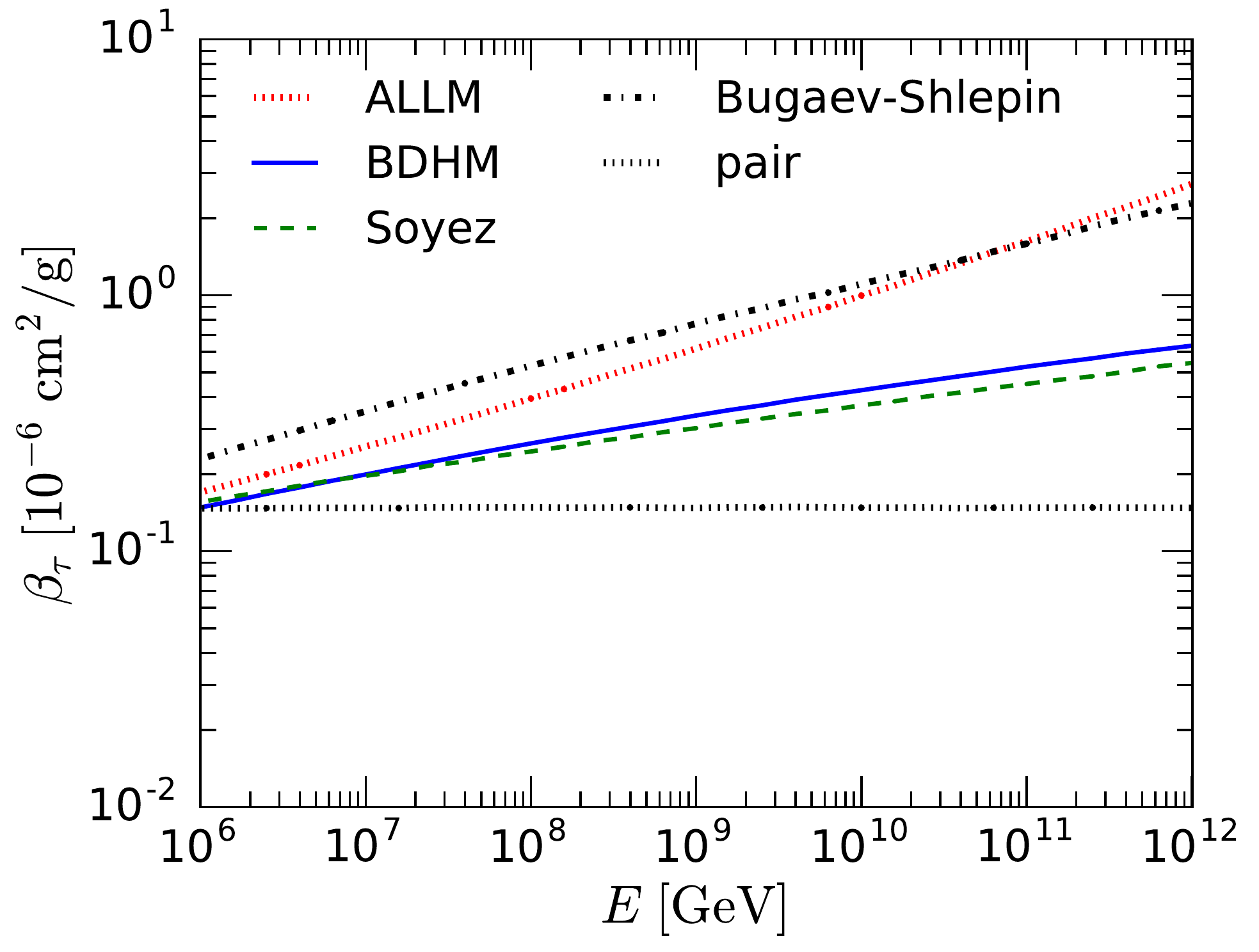}
	\caption{As in Fig. \ref{fig:betamuon}, for taus in rock. The bremsstrahlung contribution to $\beta_\tau$ is lower than the scales shown in this plot.}
	\label{fig:betatau}
\end{figure}

Our focus is on tau energy loss, however, we show the muon $\beta_\mu(E)$ to show the differences between approaches. 
In Figs. \ref{fig:betamuon} and \ref{fig:betatau}, we show the photonuclear contributions to $\beta_\ell$ for muons and taus as a function of 
energy for rock. 
For comparison, we also show the Bugaev and Shlepin evaluation of $\beta_\ell^{\rm nuc}$ \cite{Bugaev:2002gy}, an update of \cite{Bezrukov:1981ci}
with both soft and hard contributions,  and the pair production and bremsstrahlung (muons only) contributions to $\beta_\ell$.
For muons, the BDHM and Soyez evaluations of $\beta_\mu^{\rm nuc}$ fall in a similar range, while the  ALLM and
Bugaev-Shlepin results for $\beta_\mu^{\rm nuc}$ rise more quickly with energy. 
For taus, again the ALLM paramterization and Bugaev-Shlepin result for 
$\beta_\tau^{\rm nuc}$ rises 
 more steeply with energy than the
dipole model and BDHM parameterization results. 
In our discussion below, we will use the ALLM parameterization as representative of the steeper growth of $\beta_\tau^{\rm nuc}$ with tau energy.

As noted above, for the dipole model, we can evaluate the nuclear correction with a shadow factor and Glauber-Gribov (GG)
correction. 
 The shadow function or GG correction reduces the unshadowed energy loss photonuclear $\beta_\mu^{\rm nuc}$ by a factor of $0.73-0.75$ for $E=10^6-
10^{12}$ GeV. For the Soyez dipole model, we can also use the ASW prescription, which reduces $\beta_\mu^{\rm nuc}$
by a factor of about 10\%\ relative to the uncorrected value, for the same energy range. We note that the Soyez-ASW $\beta_\mu^{\rm nuc}$ coincides with the
AAMQS result using the GG correction or shadow factor. 
For taus, a similar nuclear suppression is obtained: a factor of $0.73-0.79$ for $\beta_\tau^{\rm nuc}$ with the shadow
function or GG correction, and $0.90-0.94$ for the ASW correction for the same energy range for the Soyez dipole cross section.

For taus, energy loss effects can be largely determined by $\beta_\tau$ rather than a full Monte Carlo when considering incident tau neutrino fluxes \cite{Chirkin:2004hz,Dutta:2005yt,Koehne:2013gpa}, as we discuss below. 
In Ref. \cite{Dutta:2005yt}, we found that a reasonable parameterization of $\beta_\tau$ is
\begin{equation}
\label{eq:beta2}
\beta_\tau^{\rm fit}(E) = \beta_0+\beta_1 \ln (E/E_0)\ ,
\end{equation}
for $E_0=10^{10}$ GeV and initial tau $E=10^8-10^{12}$ GeV. 
The bremsstrahlung and pair contributions in this energy range are essentially constant,
so $\beta_0 = \beta_0^{\rm brem}+\beta_0^{\rm pair}+\beta_0^{\rm nuc}$ and
$\beta_1 = \beta_1^{\rm nuc}$. Parameterizations for $\beta_\tau^{\rm nuc}$ are listed in Table \ref{table:betatau}.
In this energy range, the Soyez  and BDHM results are such that the ratio
of the numerical evaluation to the fit form gives $0.97<\beta_\tau^{\rm nuc}/\beta_\tau^{\rm fit,nuc}<1.06$. The fits are best above $E=10^9$ GeV, within $\pm 3\%$. We have also included in the Table the parameterization of $\beta_\tau^{\rm nuc}$ for the AAMQS dipole calculation of the average energy loss of the tau. The ALLM $\beta_\tau^{\rm nuc}$ is not well fit with two parameters. 
A three parameter fit to $\beta_\tau^{\rm nuc}$ is a better match to the calculated $\beta_\tau^{\rm nuc}$ for all the models, as discussed in Appendix B. For the ALLM evaluation of $\beta_\tau^{\rm nuc}$, the three parameter fit does well for $E=10^8-10^{11}$ GeV.
We focus our discussion in the text on the two parameter fit. For the flux results below, the inclusion of $\beta_2$ in the fit does not impact flux predictions except for the ALLM case. We use the three parameter fit for the ALLM case.

\begin{table}[htp]
\caption{ Fit to $\beta_\tau^{\rm nuc}$ in [$10^{-6}$ cm$^2$/g] using Eq. (\ref{eq:beta2}) for $E=10^8-10^{12}$ GeV. For
bremsstrahlung, $\beta_0^{\rm brem} =7.9\times 10^{-9}$ cm$^2$/g and pair production,
$\beta_0^{\rm pair}=0.148\times 10^{-6}$  cm$^2$/g, essentially independent of energy for this energy range.  }
\begin{center}
\begin{tabular}{|c|c|c|}
\hline
Model &$\beta_0^{\rm nuc}$ & $\beta_1$ \\
\hline 
\hline 
Soyez-Shadow & 0.380& $3.21\times 10^{-2}$ \\
\hline 
Soyez-ASW & 0.471 & $3.91\times 10^{-2}$  \\
\hline 
Soyez-GG & 0.383 & $3.07\times 10^{-2}$ \\
\hline 
AAMQS-GG& 0.433 &$3.46\times 10^{-2}$  \\
\hline 
BDHM-Shadow & 0.435 &$4.04\times 10^{-2}$  \\
\hline
\end{tabular}
\end{center}
\label{table:betatau}
\end{table}%

\subsection{Survival probability and  average range}

\begin{figure}[htb]
\centering
	\includegraphics[width=\columnwidth]{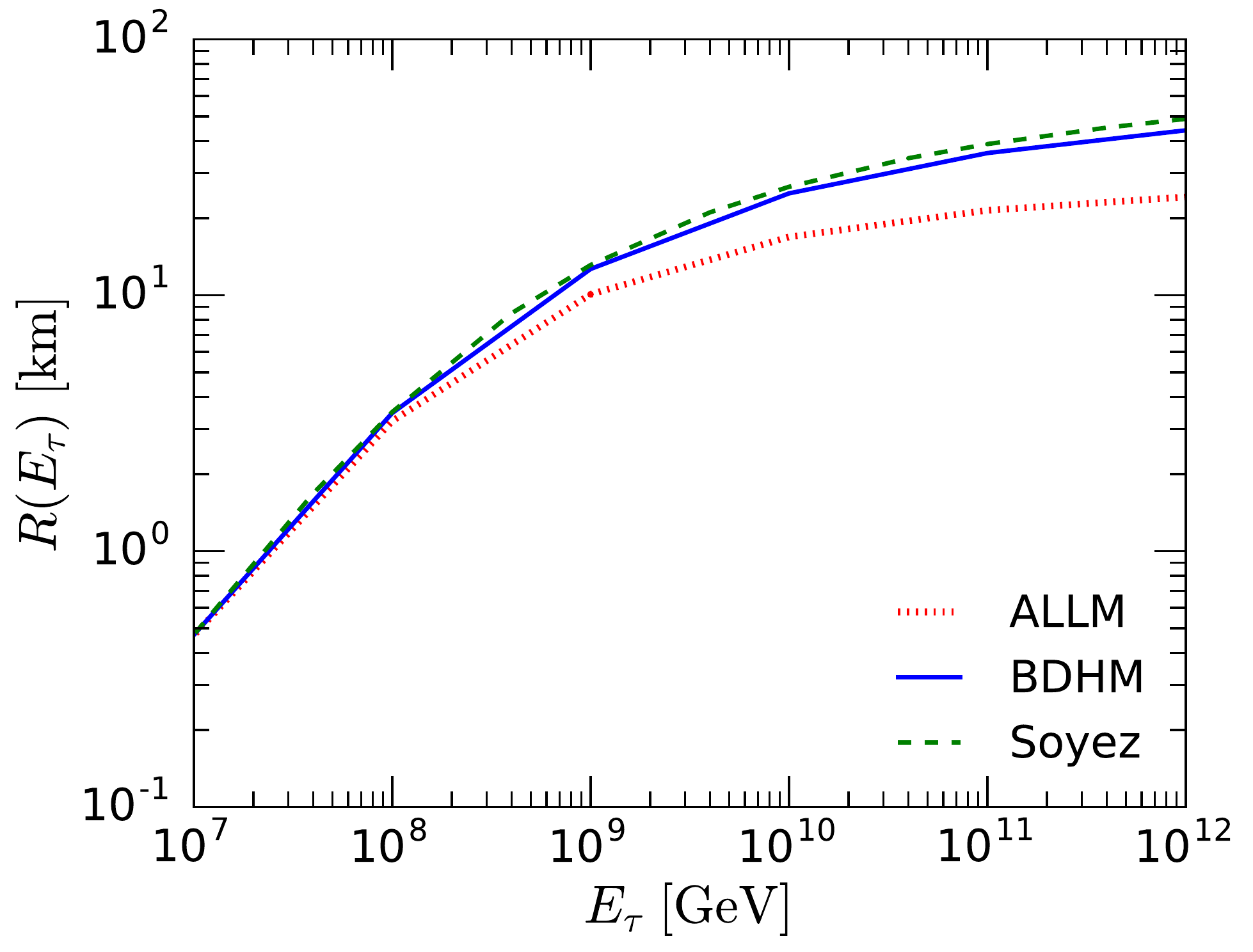}
	\caption{The average tau range in rock in km evaluated using the transport Monte Carlo with ALLM, BDHM and Soyez models for electromagnetic energy loss. Here, $E_\tau$ denotes the initial tau energy.}
	\label{fig:taurange}
\end{figure}

We start with discussion of the average lepton range to allow for 
 comparisons with other authors. 
It has been known for a long time that approximating
\begin{equation}
\label{eq:beta-approx}
\Biggl\langle\frac{dE}{dX}\Biggr\rangle\simeq \frac{dE}{dX}
\end{equation}
does not accurately represent the  average muon range because of the effect of energy fluctuations in electromagnetic interactions \cite{Lipari:1991ut}. Muons are essentially stable at these energies, however, the tau's lifetime is important here.
Following the prescription described in \cite{Lipari:1991ut} and others \cite{Antonioli:1997qw,Sokalski:2000nb,Chirkin:2004hz,Bigas:2008ff,Koehne:2013gpa}, we use a Monte Carlo program to evaluate the tau survival probability. For $y<10^{-3}$, the energy loss is evaluated in the continuous approximation, but for $y>10^{-3}$, it is evaluated stochastically. 
The average range $R(E_\tau^i)$, the average distance the tau travels, is determined by integrating the survival probability $P_{\rm surv}(E_\tau^i,X)$ as a function of column depth for taus with an initial energy $E_\tau^i$ that survive to an energy larger than $E_{\tau}^f$,
\begin{equation}
R(E_\tau^i)=\int dX\, P_{\rm surv}(E_\tau^i,X)\ .
\end{equation}
In the Monte Carlo evaluation of the average range, we use $E_{\tau}^f=50 $ GeV
for the tau, however, the short lifetime of the $\tau$ dominates its range below $E_\tau^i\sim 10^8 $ GeV. 
Our Monte Carlo evaluation of the  average tau range
is shown in Fig. \ref{fig:taurange} and in Table \ref{table:taurangerock}. 
Our results, based on our Monte Carlo developed for Ref. \cite{Dutta:2000hh} are in good agreement with 
Ref. \cite{Bigas:2008ff}. They are in qualitative agreement with those in 
 Refs. \cite{Chirkin:2004hz} and \cite{Koehne:2013gpa} where energies over many orders of magnitude are shown. For incident tau neutrino fluxes that fall with energy, the average range, without detailed final energy information, is less relevant than the survival probability as a function of final energy, as we discuss in Sec. IV.

In the approximations of
Eqs. (\ref{eq:beta2}) and (\ref{eq:beta-approx}),
just accounting for energy loss over a distance $x=X/\rho$, the initial tau energy $E_\tau^i$, final tau energy $E_\tau$ and distance traveled are related
by integrating $dE/dX$ to get
\begin{eqnarray}
\nonumber
E_\tau(E_\tau^i,x) &\simeq&\exp\Biggl[ -\frac{\beta_0}{\beta_1}\bigl( 1-e^{-\beta_1\rho x}\bigr)\\
&+& \ln(E_\tau^i/E_0)e^{-\beta_1\rho x}\Biggr] E_0
\label{eq:etau}
\end{eqnarray}
for the 2 parameter fit to $\beta_\tau(E)$. For our discussion of tau fluxes, it will be useful to also write
$E_\tau^i$ in terms of the final $E_\tau$ after travelling distance $x$:
\begin{eqnarray}
\nonumber
E_\tau^i(E_\tau,x) &\simeq&\exp\Biggl[ -\frac{\beta_0}{\beta_1}\bigl( 1-e^{\beta_1\rho x}\bigr)\\
&+& \ln(E_\tau/E_0)e^{\beta_1\rho x}\Biggr] E_0 \ .
\label{eq:etaui}
\end{eqnarray}
Appendix B shows the relation between $E_\tau$, $E_\tau^i$ and $x$ for the three parameter fit.

With the lifetime, one can approximate the survival probability \cite{Reya:2005vh,Bigas:2008ff}, 
\begin{equation}
\label{eq:psurv}
P_{\rm surv} (E_\tau^i,z)= \exp\Biggl[ -\int _0^z\frac{dx}{c\tau_\tau E_\tau(E_\tau^i,x)/m_\tau}\Biggr]\ .
\end{equation}
An explicit expression for the survival probability for the two parameter fit to $\beta_\tau(E)$, in the first order expansion in terms of $\beta_{1}/\beta_0$, 
is shown in Ref. \cite{Dutta:2005yt}. In this paper, we present the survival probability for the three parameter fit in Appendix B.
Integration of Eq. (\ref{eq:psurv}) to get an approximation to the average tau range, as in the muon case, overestimates
the average range. Nevertheless, as we discuss below, the flux of tau is well approximated by using the analytic formulas for 
the survival probability.

\begin{table}[htp]
\caption{The average tau range [km] in rock as a function of initial tau energy using our tau transport Monte Carlo.}
\begin{center}
\begin{tabular}{|c|c|c|c|c|c|}
\hline
Energy & Shadow & Shadow & Shadow  & GG\\
& BDHM & ALLM & Soyez & Soyez\\
\hline 
\hline 
 $10^8$ & 3.46 & 3.22 & 3.49 & 3.46\\
\hline 
 $10^9$ & 12.7 & 10.1 & 13.1 & 12.8\\
\hline 
 $10^{10}$ & 25.0 & 16.9 & 26.4 & 25.6\\
\hline 
 $10^{11}$ & 35.9 & 21.5 & 38.4 &37.2\\
\hline 
 $10^{12}$ & 44.1 & 24.2 & 47.2 & 45.9\\
 \hline
 \end{tabular}
 \end{center}
 \label{table:taurangerock}
 \end{table}

We remark that the
tau weak interaction energy losses are small compared electromagnetic energy loss except at the very highest 
energies. Above approximately $E=10^{10}$ GeV, tau charged current interactions are more important than tau
decays for eliminating taus from the flux, however, in this regime, electromagnetic energy loss dominates the
tau interactions. Weak neutral current contributions to the tau energy loss are quite small \cite{Kuzmin:2007zz}. We include
weak interaction effects in the Monte Carlo program for tau propagation. Without the tau charged current
interaction, the tau range is between 1.3-8\% (0.8-3.5\%) higher for $E_\tau=10^8-10^{12}$ GeV for the
evaluation of the energy loss using the BDHM (ALLM) parameterization. 

\subsection{Charged lepton spectra}

Before turning to the full process of tau neutrino conversion to taus, followed by tau propagation to exit the Earth,
we use a tau flux as a 
proxy for the neutrino flux, as was done in 
Ref. \cite{Bigas:2008ff}.  This is a useful exercise to see the features of the tau propagation without the additional element of neutrino interactions, energy transfer and neutrino flux attenuation. Our results confirm the conclusions of Ref. \cite{Dutta:2005yt,Bigas:2008ff}, namely,
that except at high energy, 
evaluations of the propagation of taus using the analytic survival probability and Eq. (\ref{eq:etaui}) reliably reproduce
the results of the full Monte Carlo propagation for an incident $(E_\tau^i)^{-2}$ flux propagating over distances of 1--20 km.

\begin{figure}[htb]
\centering
	\includegraphics[width=\columnwidth]{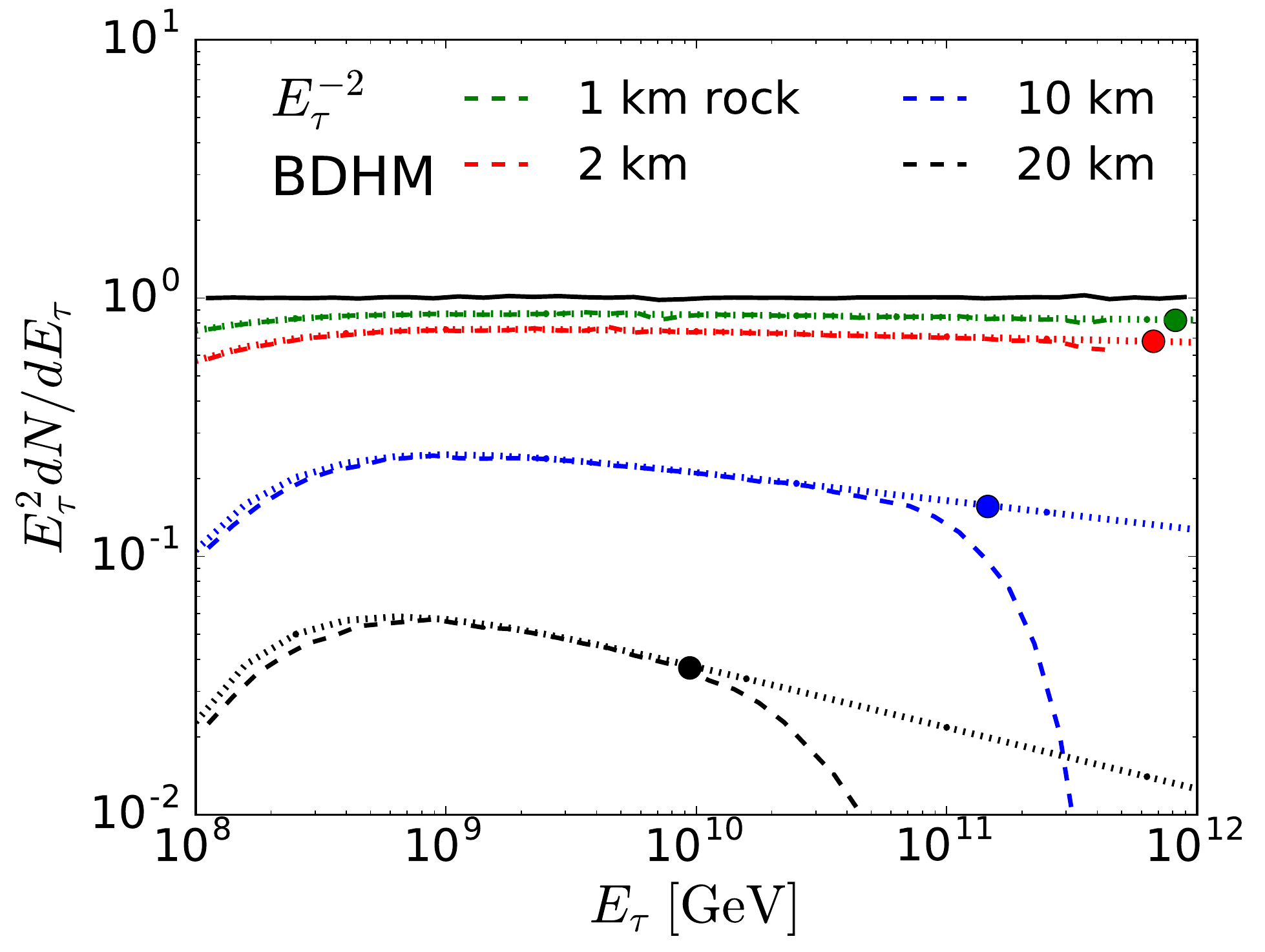}
	\caption{The number of taus per unit energy, assuming an initial flux scaling like $(E_\tau^i)^{-2}$, propagated with the Monte Carlo transport program through
	1 km, 2 km, 10 km and 20 km of rock in descending order in the figure (dashed), compared with the analytic results from Eq. (\ref{eq:tausurv}) with
	$\beta_0=0.591\times 10^{-6}$ cm$^2$/g, $\beta_1=0.0404\times 10^{-6}$ cm$^2$/g from the BDHM parameterization of the energy loss (dotted). The colored dots show the maximum $E_\tau$ for the analytic approach with $E_\tau^i<10^{12}$ GeV. The solid line shows the input tau flux to the Monte Carlo program.}
\label{fig:taufluxem2}
\end{figure}

In Fig. \ref{fig:taufluxem2}, we show the tau flux with an $(E_\tau^i)^{-2}$ incident spectrum scaled by $E_\tau^2$ (solid line), and the resulting
flux after propagation of 1 km, 2 km, 10 km and 20 km of rock with the BDHM 
parameterization of the photonuclear contribution (dotted and dashed lines). The Monte Carlo results are shown with dashed lines, which cut off at high energies compared with the analytic approximation
shown with the dotted lines.
The Monte Carlo generated an input spectrum for $E_\tau^i=10^8-10^{12}$
GeV. Our transport code was developed with look-up tables that extend to $10^{12}$ GeV, so the resulting tau flux from the Monte Carlo propagation shown here comes
from an input flux with
a sharp cutoff at $E_\tau^i=10^{12}$ GeV.

The dotted lines in the figure show the analytic formula for the propagated flux at column depth $X$:
\begin{equation}
\label{eq:tausurv}
\frac{dN}{dE_\tau} = K(E_\tau^i)^{-2}P_{\rm surv}(E_\tau^i,E_\tau)\frac{E_\tau^i}{E_\tau}
e^{\beta_1 X}\ ,
\end{equation}
where $E_\tau^i$ comes from Eq. (\ref{eq:etaui}) in the 2 parameter model for $\beta_\tau$. Here, we show the analytic result for the
$(E_\tau^i)^{-2}$ tau flux without a cutoff.  A sharp cutoff for $E_\tau^i$ translates directly to a sharp cutoff in $E_\tau$ for a fixed distance. The colored dots in Fig. \ref{fig:taufluxem2} show $E_\tau^{\rm max}(E_\tau^{i\,{\rm max}},x)$ for $x=1,\ 2,\ 10$ and 20 km in rock
for a sharp cutoff at $E_\tau^{i\, {\rm max}}=10^{12}$ GeV, 
the same cutoff used in the Monte Carlo. If one would impose the same 
sharp cutoff in analytic evaluation as was done in our 
MC, the tau flux would have 
a steep drop for energies above those labeled with the colored dots.
For $E_\tau>E_{\tau}^{\rm max}$, $E_\tau^2 dN/dE_\tau\to 0$ in the analytic evaluation with a cutoff for $E_\tau^i$.

Fig. \ref{fig:taufluxem2}  shows the role of the lifetime, which suppresses the lowest energy. With increasing depth, the flux becomes steeper.
A priori, the agreement between the analytic and the Monte Carlo results is not obvious for energies between where the lifetime does not dominate and sufficiently lower than shown by the colored dots in Fig. \ref{fig:taufluxem2}. The Monte Carlo results include energy fluctuations, and as noted above, the average range evaluated from the Monte Carlo
is shorter than what would be calculated by integrating the survival probability.

The agreement between the analytic and Monte Carlo evaluations comes from competing effects, including the fact that the flux decreases with energy.
For a fixed incident tau energy and propagation distance, the Monte Carlo transport yields a slightly higher average final tau energy 
than the analytic final energy 
evaluated with Eq. (\ref{eq:etau}). Fig. \ref{fig:taufluxem2} shows the flux as a function of the final tau energy, so instead, the comparison should be of the initial tau energy. The average corresponding incident tau energy in the Monte Carlo approach is lower than the corresponding analytic value (see Eq. (\ref{eq:etaui})). The survival probability as evaluated by the Monte Carlo approach is also lower than from the analytic formula in
Eq. (\ref{eq:psurv}). Schematically, Eq. (\ref{eq:tausurv}) shows that decreases in the incident tau energy and survival probability will tend to compensate. Our numerical comparison shows that this is the case for the $(E_\tau^i)^{-2}$ incident flux.
Thus comparison of the tau flux (i.e. final tau energy) in the two approaches
effectively corresponds to different values of the $E_\tau^i$ and somewhat different survival probabilities.

To evaluate the tau flux due to incident neutrinos passing through a depth of rock, we need the neutrino cross section, and the differential cross section as a function of $y$. In the next section we show results for neutrino cross sections as evaluated by different models, then use the Monte Carlo to propagate the neutrinos and produced taus through rock.

\section{Neutrino cross sections}

The flux of taus coming from tau neutrino conversions in the Earth depends on the neutrino cross section, both for
neutrino attenuation and for neutrino conversion to taus, and on the differential distribution as a function of inelasticity $y$.
As shown in Section II, different models for the electromagnetic structure function $F_2$ have different low-$x$ predictions. 
The neutrino cross sections at high energies probe the low-$x$ behavior of the weak structure functions, albeit at a higher value of $Q^2$, namely $Q^2\sim M_W^2$ (see, e.g., Refs. \cite{Gandhi:1998ri,Reno:2004cx}). 

\begin{figure}[htb]
\centering
	\includegraphics[width=\columnwidth]{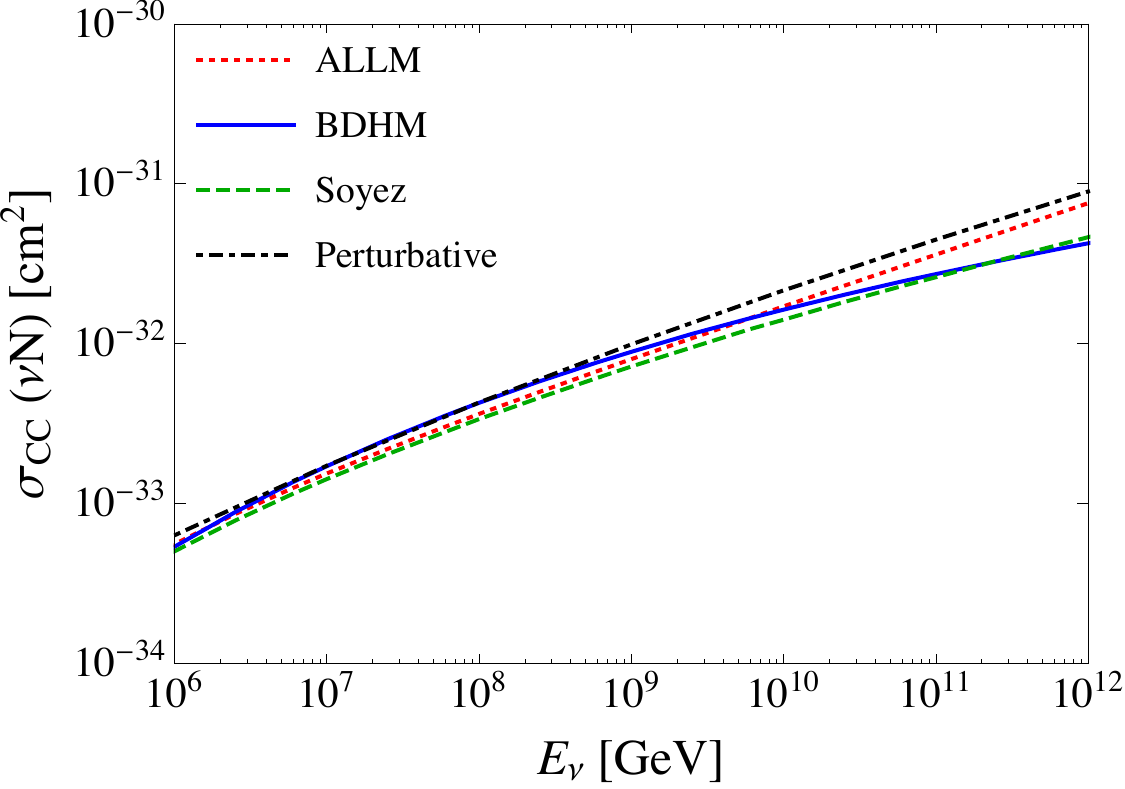}
	\caption{The neutrino nucleon charged current interaction $\sigma_{\nu N}$ as a function of neutrino energy $E_\nu$ from NLO perturbative QCD \cite{Jeong:2010za} (dot-dashed), using the Soyez dipole (dashed) and the BDHM (solid) and ALLM (dotted) parameterizations.}
	\label{fig:sigcc}
\end{figure}

For a fixed column depth, considering both the probability for neutrino interactions to produce the tau lepton, and
the tau lepton energy loss, there are competing effects due to the low-$x$ behavior of the structure functions. For the ALLM parameterization, the neutrino cross section is enhanced relative to the BDHM parameterization result, 
as shown in Fig. \ref{fig:sigcc} with the dotted and solid lines. However, there is also
a greater energy loss of the tau at high energies because of the same small-$x$ enhancement. For the results below,
we keep the same model for both the neutrino interaction and the tau interaction.

Detailed formulae for the neutrino weak interactions are shown in Appendix C. For the dipole model, the
wave functions for $W$ or $Z$ fluctuations to a quark-antiquark pair are functions of the quark masses 
\cite{Fiore:2011gx,Barone:1993es,Kutak:2003bd,Arguelles:2015wba}. We include them in Appendix C, along with the relevant couplings.
 Given, e.g., the Soyez dipole cross section,
the ultrahigh energy neutrino cross section is predicted, as shown in Fig. \ref{fig:sigcc} with the (green) dashed curve.  

For $E>10^7$ GeV,
valence contributions, neglected here, are small.  Fig. \ref{fig:sigcc}
 shows that these three approaches yield nearly the same
neutrino-nucleon cross sections at $E_\nu=10^6-10^7$ GeV. As noted in Ref. \cite{Armesto:2007tg}, the ALLM electromagnetic structure function lies somewhat below the HERA data for $Q^2\sim 200-8000$ GeV$^2$ and $x >10^{-2}$. We note that measurements of the neutral current structure function at $Q\sim M_W$ are available only for $x>0.1$ 
\cite{Aaron:2009aa}, the scale relevant to the ultrahigh energy neutrino cross section \cite{Gandhi:1998ri}. 
The ALLM parameterization yields neutrino cross sections that are too low below $E_\nu\sim 10^5$ GeV. Since we focus here on energies even higher than $10^6$ GeV, we are probing very small-$x$ values where the ALLM parameterization is useful as a contrast to other approaches with its stronger dependence on $x$. 

The explicit inclusion of quark channels in the dipole model through the wave function squared cannot be done with the electromagnetic structure function 
parameterizations of ALLM and BDHM. Instead, we rescale the electromagnetic structure function
by the sum of electric charges squared for five quark flavors, to yield
\begin{eqnarray}
F_2^{CC}&\simeq& \frac{45}{11} F_2^{\gamma} \\
F_2^{NC}&\simeq&\biggl(  \frac{45}{22}-\frac{41}{11}\sin^2\theta_W+4\sin^4\theta_W\biggr)
 F_2^{\gamma} \ .
\end{eqnarray}
We use the Callan-Gross relation $2xF_1=F_2$ and set $F_3=0$ for neutrino interactions when we use the
BDHM and ALLM parameterizations of $F_2$. The charged current cross sections for these models are also
shown in Fig. \ref{fig:sigcc}, along with the next-to-leading order perturbative evaluation of the neutrino charged-current
cross section from Ref. \cite{Jeong:2010za}. At $E_\nu=10^6$ GeV, the cross sections are within $\pm 5\%$ of each other,
increasing to $\pm 12\%$ for $E_\nu=10^8$ GeV. As the the neutrino energy further increases, the spread in cross sections increase to a factor of $\sim 1.7$ for the BDHM and Soyez approach compared to the ALLM extrapolation at $10^{12}$ GeV. For reference, recent evaluations of the neutrino cross sections with a broader range of models \cite{CooperSarkar:2011pa,Fiore:2011gx,Goncalves:2015fua,Albacete:2015zra} show a similar spread in cross sections at these energies.

\begin{figure}[h]
\centering
	\includegraphics[width= \columnwidth]{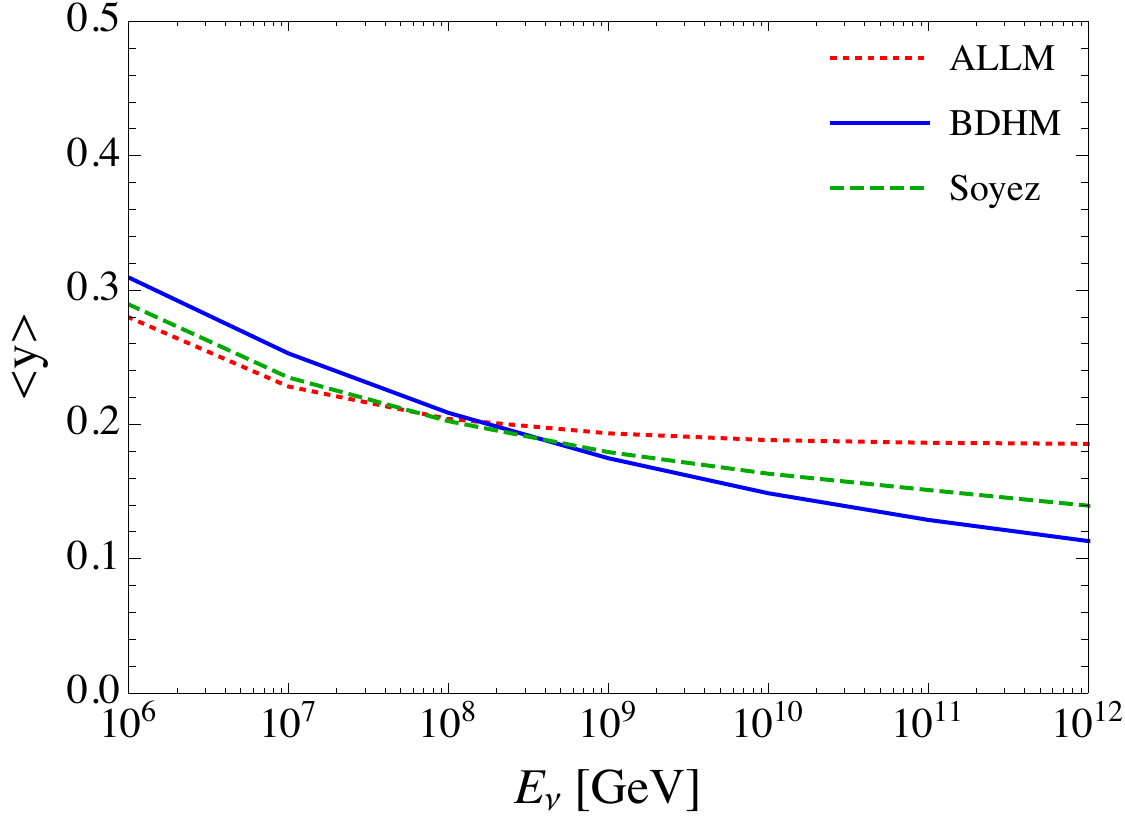}
	\caption{The mean inelasticity $\langle y\rangle$ for neutrino scattering with isoscalar nucleons as a function $E_\nu$ using the Soyez dipole (dotted) and parameterizations of BDHM (dashed) and ALLM (solid).}
	\label{fig:meany}
\end{figure}

The tau flux also requires a knowledge of the energy transferred from the neutrino to the tau, 
namely, the distribution as a function of 
$y\equiv (E_\nu-E_\tau)/E_\nu$. The average $y$ as a function of 
energy is shown in Fig. \ref{fig:meany}. 
We note that the average $y$ varies from 
$-3\%$ to $ 7\%$ at $10^6$ GeV and $-15\%$ to $23\%$ at $10^{11}$ GeV 
with respect to the values from the Soyez approach.  This range of the 
average $y$ represents theoretical uncertainty in the elasticity due to different approaches that we consider. 
In the Monte Carlo evaluation, we use look-up tables for the neutrino cross section
and distribution in $y$ up to a maximum $E_\nu=10^{12}$ GeV, however, we also do a semi-analytic evaluation
using $\langle y\rangle$.
We show in the next section that $\langle y\rangle$ and a parameterization of the neutrino cross section works well
to reproduce the Monte Carlo results. We include here a parameterization of the neutrino charged current cross section and average $y$. For BDHM and Soyez, we find
\begin{equation}
\label{eq:sigfit}
\sigma (E) =  [\sigma_0 + \sigma_1 \ln (E) + \sigma_2 \ln^2(E) ]\cdot 10^{-31} {\rm cm^2}\ ,
\end{equation}
while for ALLM and a NLO perturbative evaluation (JR in Ref. \cite{Jeong:2010za}),
\begin{equation}
\label{eq:sigfit-a}
\sigma (E) =\sigma_{0}   E^{\sigma_{1}}  \cdot 10^{-35} {\rm cm^2} \ ,
\end{equation}
for $E$ in GeV.
The average $y$ can be written as
\begin{equation}
\label{eq:meany}
\langle y (E) \rangle  = y_0 + y_1 \ln (E) + y_2 \ln^2(E)\ .
\end{equation}
The parameters $\sigma_i$ and $y_i$ are shown 
in Tables \ref{table:sigcc} and \ref{table:avgy}. 
The cross section parameterizations reproduce the charged current 
cross sections to within $\pm 5\%$. The parameterizations reproduce 
$\langle y\rangle$ to within $\pm 1\%$.

\begin{table}[h]
\caption{Parameters for the charged current neutrino cross sections from Eq. (\ref{eq:sigfit}) and  (\ref{eq:sigfit-a}).}
\begin{center}
\begin{tabular}{|c|c|c|c|c|}
\hline
Model & E [GeV] & $\sigma_0$ & $\sigma_1$ & $\sigma_2$\\
\hline 
\multirow{2}{*}{BDHM}& $10^{6} \leq E \leq 10^{9}$ & 0.310 &  $-4.45\cdot 10^{-2} $ & 1.63 $\cdot 10^{-3}$\\
\cline{2-5}
& $ E  >10^{9}$ & 1.32 & $ -0.141$ & 3.92 $\cdot 10^{-3}$\\
\hline 
\multirow{2}{*}{Soyez}& $10^{6} \leq E \leq 10^{9}$  & 0.269 & $ -3.80 \cdot 10^{-2}$ & 1.37 $\cdot 10^{-3}$\\
\cline{2-5}
& $ E  >10^{9} $ & 2.30 & $ -0.230$ & 5.91 $\cdot 10^{-3}$\\
\hline 
\multirow{2}{*}{ALLM}& $10^{6} \leq E \leq 10^{9}$  & 0.460 &  0.361 & $-$ \\
\cline{2-5}
& $E  >10^{9}   $ & 1.05 & 0.321 & $-$\\
\hline 
\multirow{2}{*}{JR \cite{Jeong:2010za}}& $10^{8} \leq E \leq 10^{10}$  & 0.718 &  0.348 & $-$ \\
\cline{2-5}
& $E  >10^{10}   $ & 1.78 & 0.308 & $-$\\
\hline
\end{tabular}
\end{center}
\label{table:sigcc}
\end{table}%

\begin{table}[h]
\caption{Parameterizations of $\langle y \rangle$ using Eq. (\ref{eq:meany}). 
The values of $\langle y \rangle$ from these parameterizations 
have less than 1\% error between $E=10^6-10^{13}$ GeV. }
\begin{center}
\begin{tabular}{|c|c|c|c|c|}
\hline
Model & E [GeV] & $y_0$ & $y_1$ & $y_2$\\
\hline 
\multirow{2}{*}{BDHM}& $10^{6} \leq E \leq 10^{8}$ & 0.909 & $ -5.95 \cdot 10^{-2}$ & 1.17 $\cdot 10^{-3}$\\
\cline{2-5}
& $10^{8} < E \leq 10^{13}$ & 0.654 & $ -3.35\cdot 10^{-2}$ & 5.01 $\cdot 10^{-4}$\\
\hline 
\multirow{2}{*}{Soyez}& $10^{6} \leq E \leq 10^{8}$  & 1.08 & $ -8.55 \cdot 10^{-2}$ & 2.07 $\cdot 10^{-3}$\\
\cline{2-5}
& $10^{8} < E \leq 10^{13}$ & 0.478 & $ -2.05 \cdot 10^{-2}$ & 2.98 $\cdot 10^{-4}$\\
\hline 
\multirow{2}{*}{ALLM}& $10^{6} \leq E \leq 10^{8}$  & 1.17 & $ -9.99 \cdot 10^{-2}$ & 2.59 $\cdot 10^{-3}$\\
\cline{2-5}
& $10^{8} < E \leq 10^{13}$ & 0.356 & $ -1.25 \cdot 10^{-2}$ & 2.27 $\cdot 10^{-4}$\\
\hline 
\end{tabular}
\end{center}
\label{table:avgy}
\end{table}%

In evaluating the tau neutrino induced tau flux, we also need the neutrino neutral current cross section for the attenuation
factor. For the range of $E_\nu=10^8-10^{12}$ GeV, the ratio of the neutral current to charged current cross section ranges between 0.42-0.44 for the BDHM 
parameterization, 0.44-0.45 for the Soyez dipole, and 0.41-0.42 for the ALLM parameterization.

\section{Results}

We begin this section by comparing a Monte Carlo evaluation of neutrino interactions followed by tau propagation through rock of various depths. 
To quantify the effects of the neutrino cross section to produce taus, attenuation of the neutrino flux and tau energy loss, we plot the ratio
of the emerging tau flux to the incoming neutrino flux -- the transmission function $F(\tau)/F(\nu)$ -- where
$$F(i) = \frac{dN}{dE_i}$$
is the number of leptons per unit energy per unit area per sec, assuming an isotropic incident flux. We only show results for incident neutrinos. For $E_\nu$ larger than $\sim 10^7$ GeV,  $\sigma_{\nu N}\simeq \sigma_{\bar{\nu}N}$ to a very good
approximation. 

In the next section, we consider distances less than 10 km, relevant to using mountains as neutrino converters. In Section IV.B, we
evaluate the transmission functions for larger distances 
where neutrino attenuation effects come into play.

\subsection{Distances less than 10 km}

\begin{figure}[htb]
\centering
	\includegraphics[width=0.895 \columnwidth]{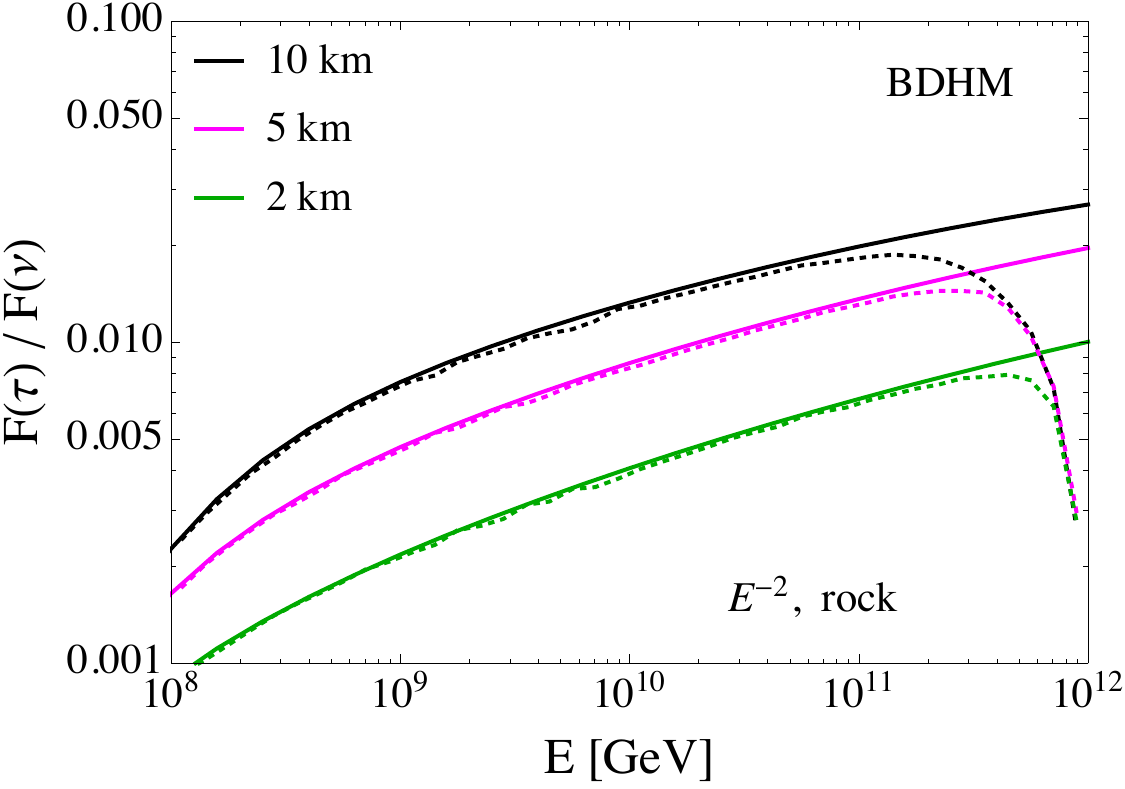}
	\includegraphics[width=0.895\columnwidth]{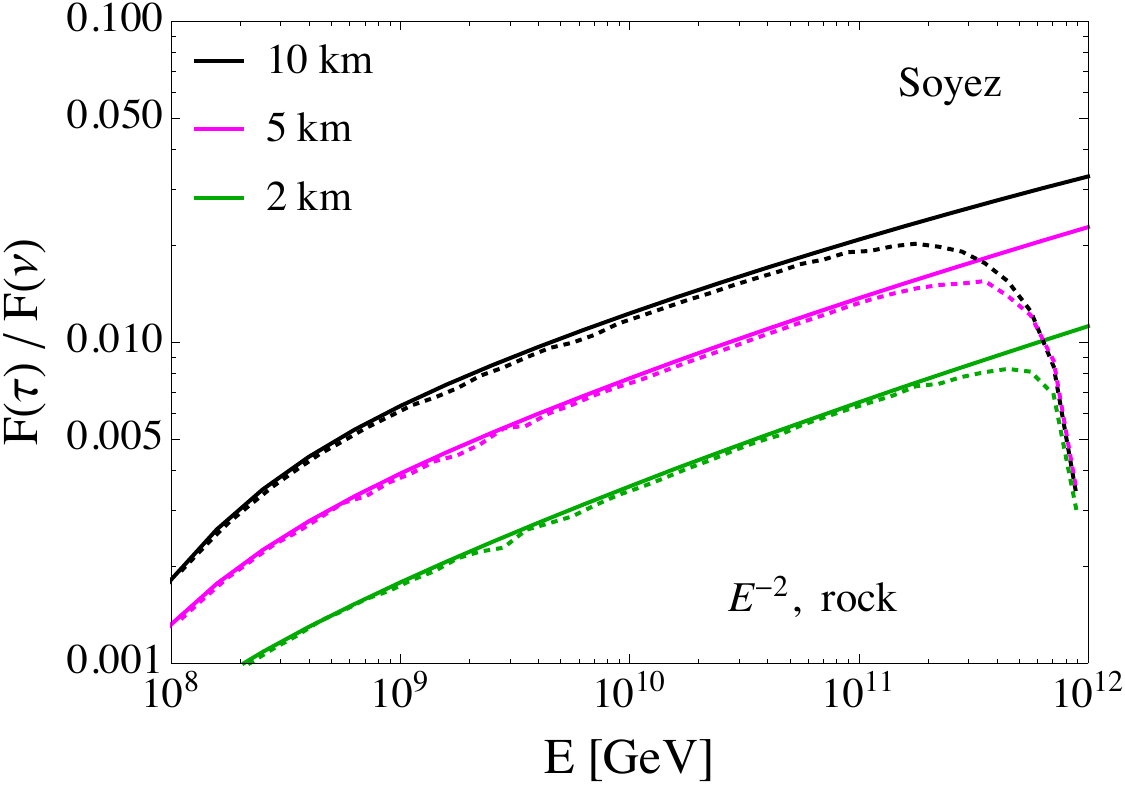}
	\includegraphics[width=0.895 \columnwidth]{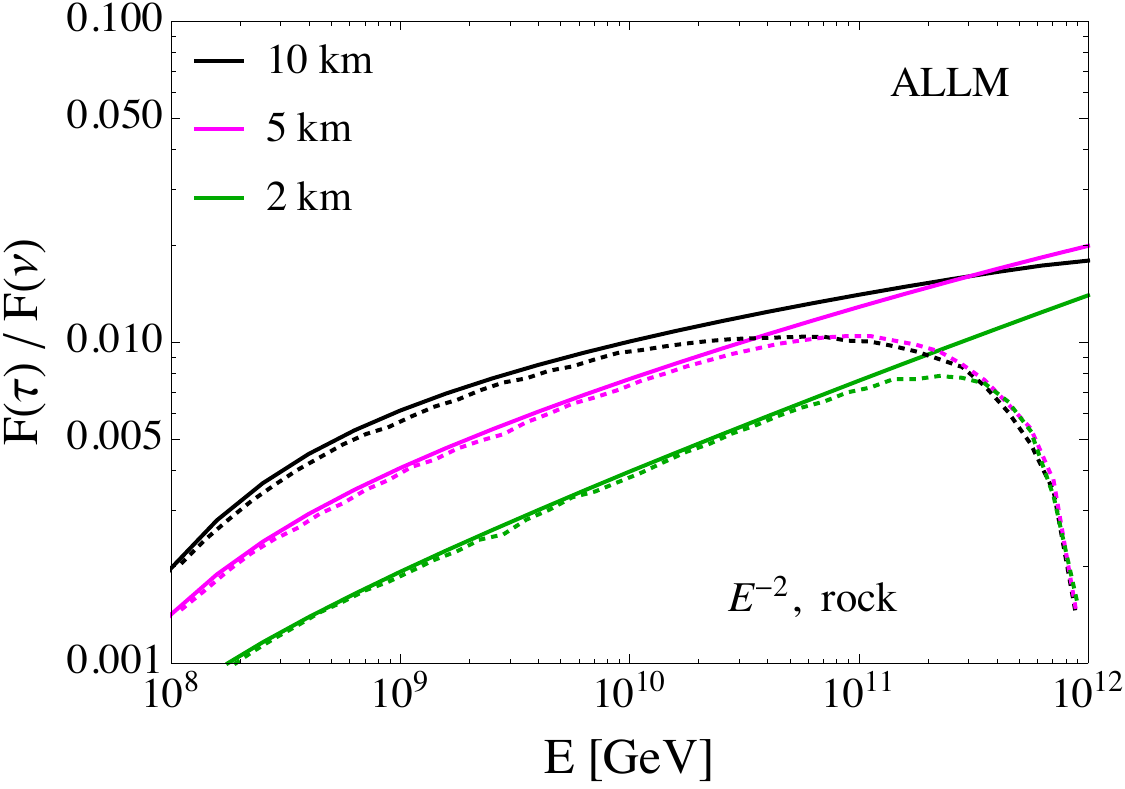}	
	\caption{The ratio of outgoing tau flux to incident tau neutrino flux with an incident flux that scales as $E_\nu^{-2}$ for a depth of 2, 5 and10 km rock, as a function of lepton energy, using the BDHM (upper), Soyez (middle) and ALLM (lower) approaches to both tau energy loss and the tau neutrino cross section. The solid lines use the analytic survival probability and the numerical differential neutrino cross section. The dashed lines use a Monte Carlo evaluation of the neutrino interaction and the tau energy loss, with a tau
neutrino energy cutoff of $E_\nu=10^{12}$ GeV.}
\label{fig:flux-diffuse-mc}
\end{figure}

In Fig. \ref{fig:flux-diffuse-mc}, we present the results from the analytic evaluation and Monte Carlo evaluation
for the depths of 2, 5,  and 10 km of the standard rock ($\rho = 2.65 \ {\rm g/cm^3}$) for all models considered here.  
The incident neutrino flux is assumed to have $E_\nu^{-2}$ spectrum. The analytic calculation uses the survival probability for the tau lepton (Eq. (\ref{eq:psurv-3par})) for the three parameter fit to $\beta_\tau(E)$ written in Eq. (\ref{eq:beta-3par}). The
tau flux, for depth D, is
\begin{eqnarray}
\nonumber
F_\tau(E_\tau) &=& \int_0^D dz \int dy \exp(-z \sigma_{tot} (E_\nu) \rho N_A) F_\nu(E_\nu,0)  \\
\nonumber
& \times &  N_A \rho  \frac{d \sigma_{CC} (E_\nu,y)}{dy} \frac{E_\nu}{E_\tau^{i \prime}} P_{surv}(E_\tau,E_\tau^{i \prime}) \\
& \times &\delta(E_\tau-E_\tau(E_\tau^{i \prime},D-z) ) dE_\tau^{i \prime}\ ,
\label{eq:ftau-results}
\end{eqnarray} 
where $E_\nu = E_\tau^i/(1-y)$.
While evaluating the analytic fluxes, we extended the neutrino cross sections 
for the energies above $10^{12}$ GeV, while
for the Monte Carlo results, the neutrino flux is cut off at $10^{12}$ GeV, leading
to tau fluxes that start to steeply decrease
at energies that depend on the depth, as shown in Fig. \ref{fig:flux-diffuse-mc}. 

Monte Carlo results are well matched to the analytic calculations for energies below the maximum tau energy in the analytic
approach. There are
less than $10 \%$ errors for $E \leq 10^{11}$ GeV for the BDHM and the Soyez dipoles, where the impact of the
neutrino energy cutoff at $10^{12}$ GeV starts to become evident in the Monte Carlo evaluation. 
For the ALLM parameterization, the similar difference appears between the Monte Carlo and analytic
results at $10^{10}$ GeV. The lower energy of the onset of the cutoff for the ALLM parameterization comes from the higher level of electromagnetic energy loss in this model. Since the Monte Carlo results are so well represented by the 
analytic approximation below $E\sim 10^{10}-10^{11}$ GeV depending on the depth, we do the rest of our analysis using the analytic tau survival probability integrated with
the neutrino differential cross section and the attenuation factor. 
By comparing the Monte Carlo results (dashed curves) with the analytic curves with no neutrino energy cutoff (solid lines), Fig. \ref{fig:flux-diffuse-mc} clearly shows a feed-down effect of the neutrino energy cutoff at $E_\nu=10^{12}$ GeV. At these short distances, the cutoff
effect will scale with the cutoff energy. In the remaining figures in Sec. IV.A, we use the analytic approximation without a cutoff for the
evaluation of the transmission function.

In the previous section, we presented the parameterizations of $\langle y \rangle$ 
and the neutrino charged current cross sections. For reference, we compare the resulting tau fluxes
using $\langle y(E)\rangle$ and the parameterized cross section with the numerical evaluation of the neutrino
differential cross sections. We evaluate $\langle y (E)\rangle $ using $E=E_\tau$.
The tau fluxes with the parameterizations deviate from the fully evaluated fluxes by at most $4 \%$ using Eq. (\ref{eq:meany}) for 
$\langle y \rangle$ and  by at most $ \pm 5 \%$ using both the parameterized $\langle y(E)\rangle$ and $\sigma_{\nu N}(E)$  
between $E =10^7 - 10^{12} \ {\rm GeV}$.

\begin{figure}[htb]
\centering
	\includegraphics[width=\columnwidth]{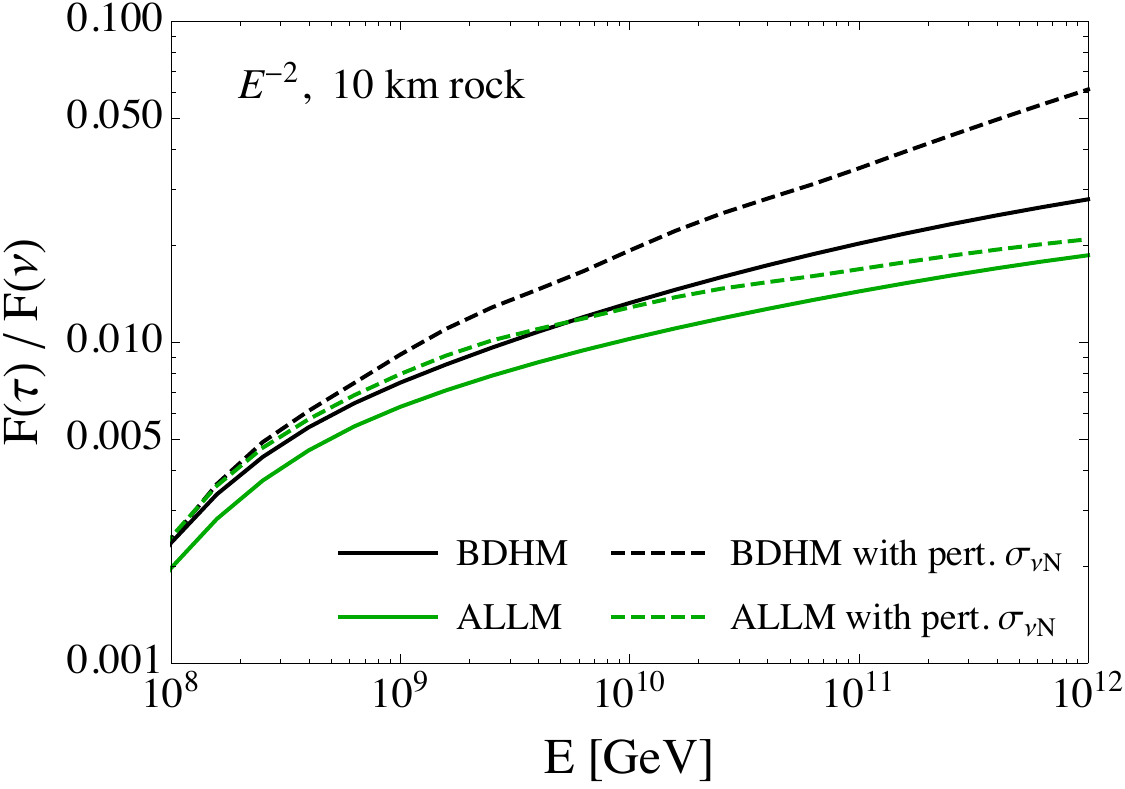}
	\caption{The ratio of outgoing tau flux to incident tau neutrino flux for a depth of 10 km rock, as a function of lepton energy, using the BDHM (upper) and ALLM (lower) tau energy loss parameterizations. The solid lines use neutrino nucleon cross sections based on the same BDHM and ALLM parameterizations, while the dashed lines show the flux ratio if the perturbative NLO neutrino nucleon cross section is used.}
	\label{fig:sigJR}
\end{figure}

One of the features of the analysis presented here is that the neutrino cross section evaluations 
use the same form for the small-$x$ 
behavior as the evaluation of the electromagnetic energy loss. At fixed depth, one can see evidence of this effect in 
Fig. \ref{fig:sigJR}. The solid lines in the figure show the ratio of the outgoing tau flux to the incoming tau neutrino flux when the parameterization (BDHM for upper curves and ALLM for lower curves) is used for both effects. The dashed curves have the BDHM or ALLM parameterizations used for the electromagnetic energy loss, but where the neutrino cross section is evaluated using NLO perturbative QCD \cite{Jeong:2010za}. The impact of the perturbative QCD choice is significant in the flux ratio at high energies when combined with the BDHM parameterization, since the small-$x$ behaviors are so different. The ALLM electromagnetic energy loss with the perturbative neutrino cross section shows less variation with a complete ALLM evaluation because the ALLM structure function and parton distribution function small-$x$ extrapolations are similar.  
We discuss the impact of using the perturbative QCD neutrino nucleon cross section below, however, for the remainder of this section, we keep consistent the 
approach to the small-$x$ behavior for both tau energy loss and neutrino nucleon scattering.

\begin{figure}[htb]
\centering
	\includegraphics[width=0.9 \columnwidth]{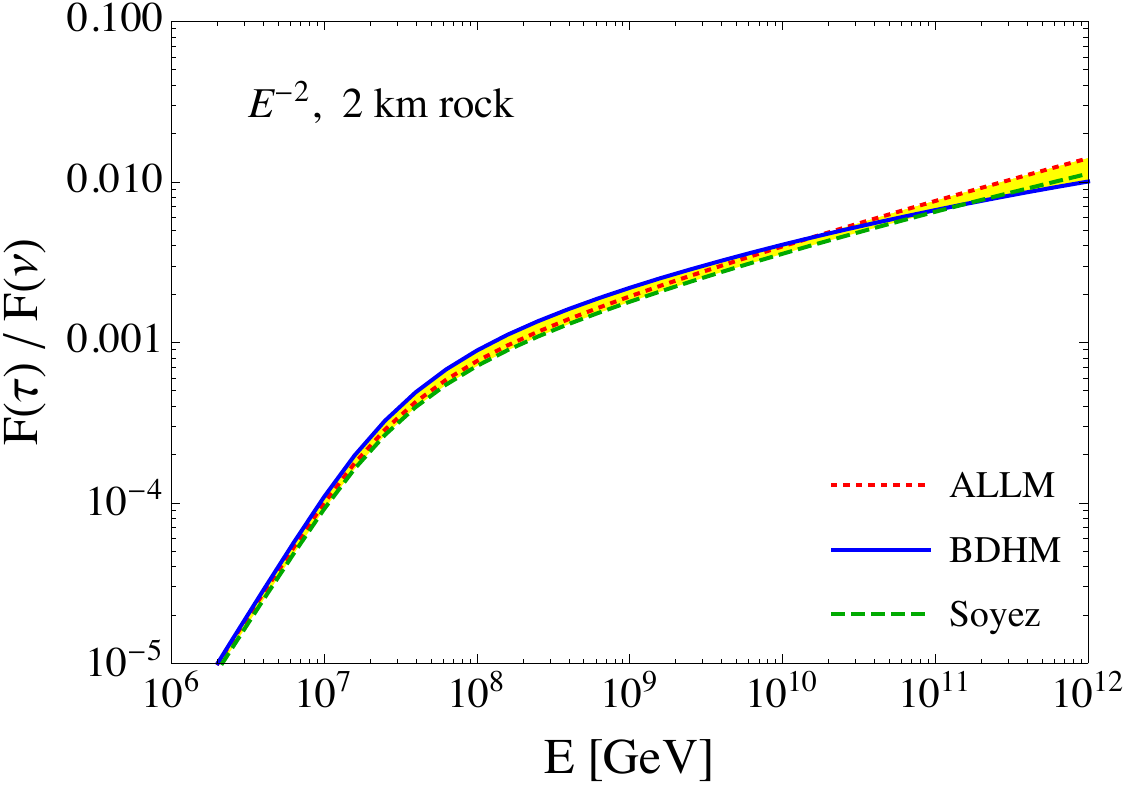}
	\includegraphics[width=0.9 \columnwidth]{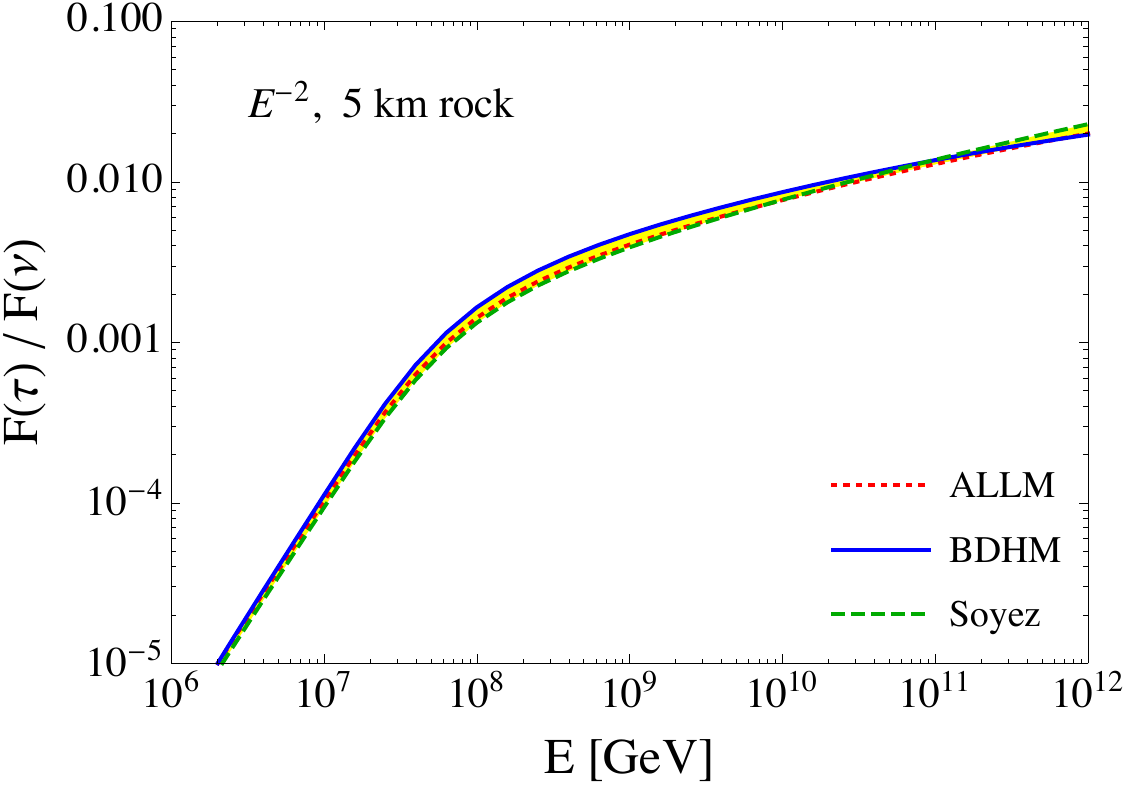}
	\includegraphics[width=0.9 \columnwidth]{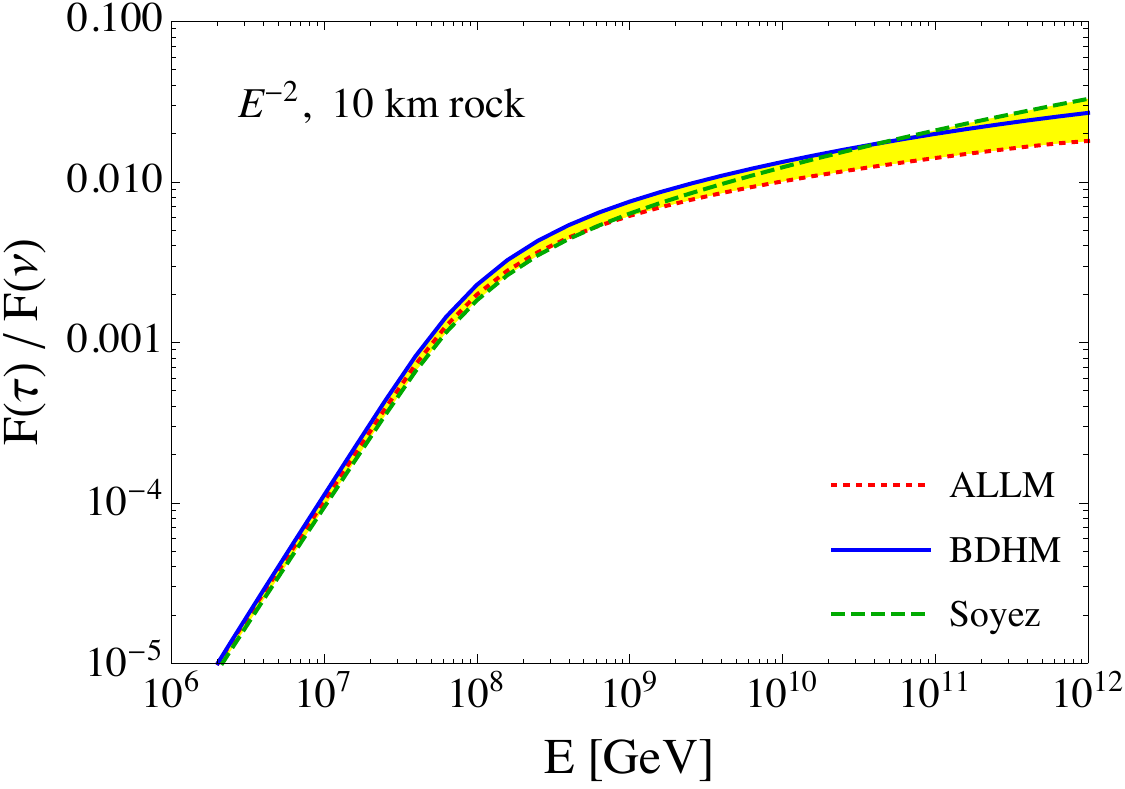}	
	\caption{The ratio of outgoing tau flux to incident tau neutrino flux for 
depths of $2$, $5$ and $10$  km rock, as a function of lepton energy, using the BDHM, Soyez and ALLM parameterizations and an incident neutrino flux scaling as $E_\nu^{-2}$.}
\label{fig:flux-diffuse-le}
\end{figure}

Fig. \ref{fig:flux-diffuse-le} compares the flux ratio from the different approaches for the various depths. 
The HAWC \cite{Vargas:2016hcp} and Ashra \cite{Asaoka:2012em} configurations are sensitive to 
Earth skimming tau neutrinos at lower energies than the Pierre Auger Observatory. 
For example, the tau neutrino energy range for the proposed Ashra experiment is PeV--EeV.
Therefore, we extend our results to lower energies, i.e., $10^6$ GeV, 
 for those experiments.
In the low energy range, below $10^8$ GeV, the tau decay process dominates, 
and the fluxes are hardly affected by the energy loss. 
Therefore, the difference of the fluxes in this energy range 
reflects mostly the ratio of the neutrino cross sections from the 
different models.  
%

As presented in Fig. \ref{fig:flux-diffuse-le}, the BDHM and the Soyez dipoles give the comparable results all over the energy range. There is at most a  25\% difference, which is its maximum at $E=10^8$ GeV.
The wider error band from all the different models depends on the behavior of the fluxes from the ALLM model compared to the BDHM/Soyez approaches. 
For the Soyez dipole, we only show the results with nuclear shadowing included by a multiplicative factor, the same factor also used with the BDHM and ALLM models. 
It is almost the same as the predicted flux ratio with GG nuclear effects.
The ASW nuclear corrections for the Soyez flux evaluation yield results at most 10 \% higher.

As shown in Fig. \ref{fig:betatau}, the energy loss from the ALLM parameterization increases more quickly with energy than the other models.
This reduces the tau flux at high energies with a higher rate, and its effect appears more significantly at larger depth.  
Thus, the uncertainty range is the largest at $10^{12}$ GeV for the $10$  km depth, 
where the flux from the Soyez dipole is the largest. It is $83 \%$ 
 higher than that from the ALLM parameterization. 
For shorter depths, tau energy loss is less important, so the higher neutrino cross 
section from the ALLM parameterization shown in Fig. \ref{fig:sigcc} results 
in the higher flux ratio for the ALLM evaluation.

\begin{figure}[htb]
\centering
	\includegraphics[width=0.9 \columnwidth]{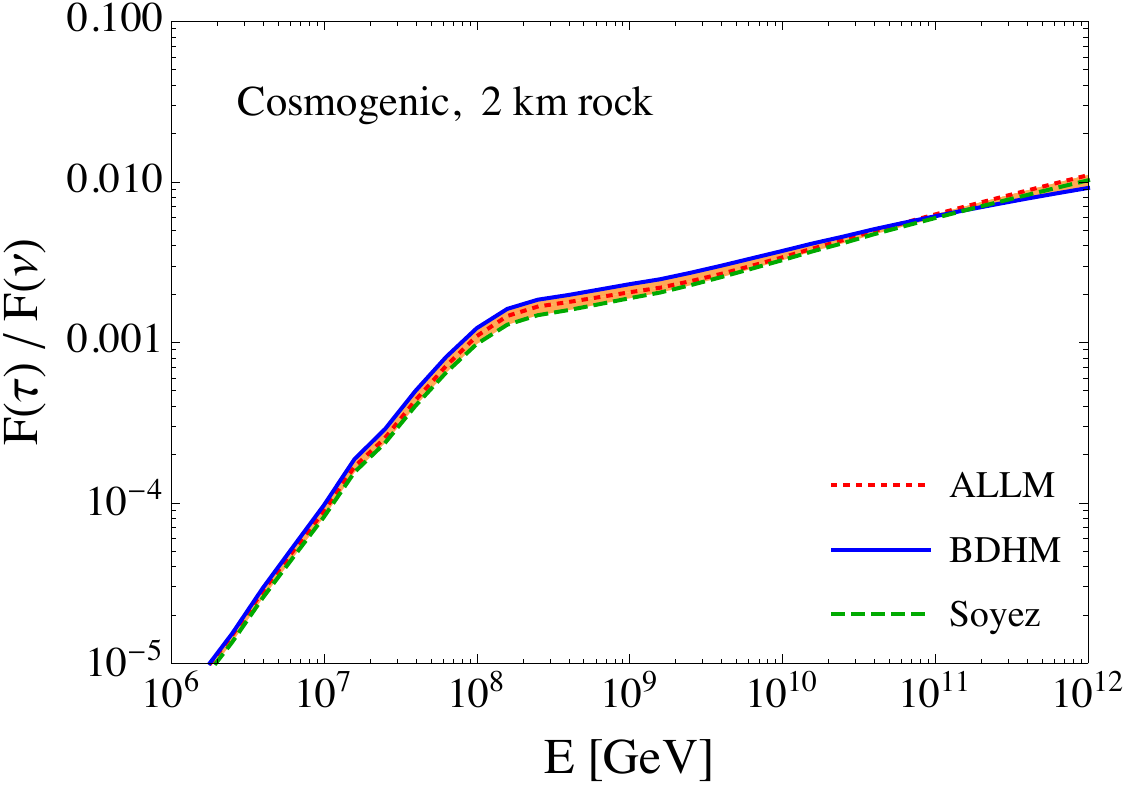}
	\includegraphics[width=0.9 \columnwidth]{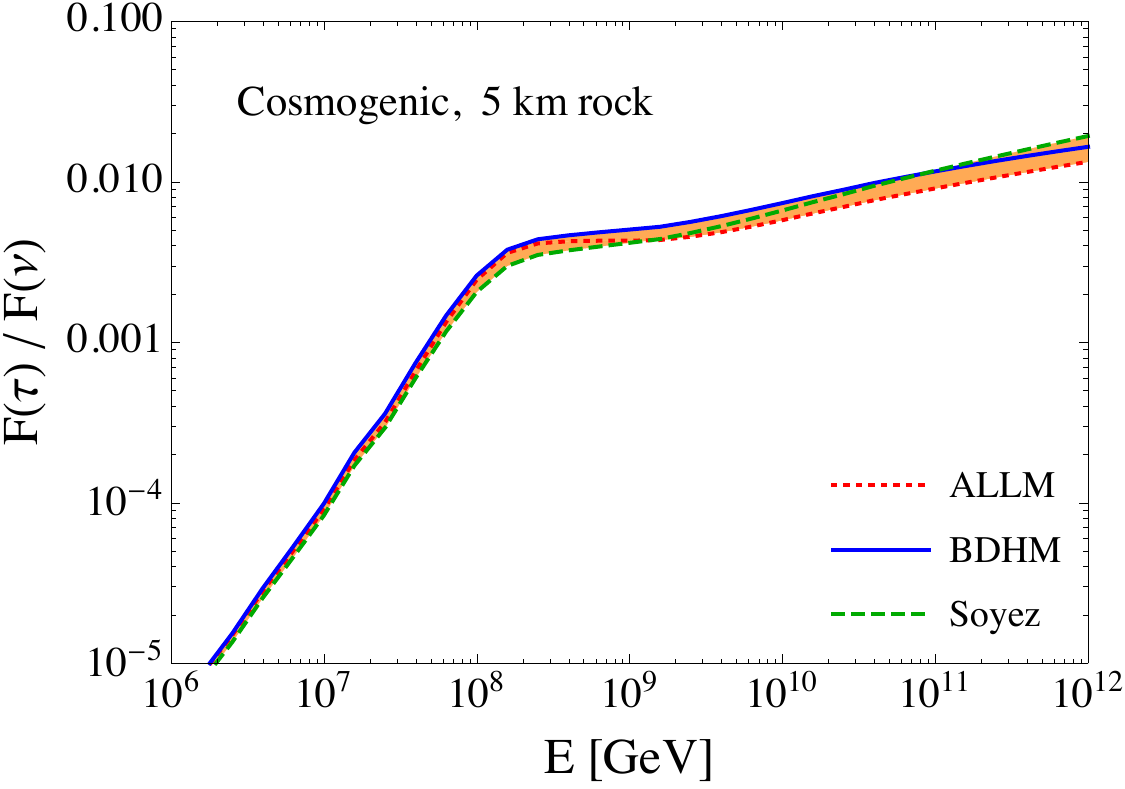}
	\includegraphics[width=0.9 \columnwidth]{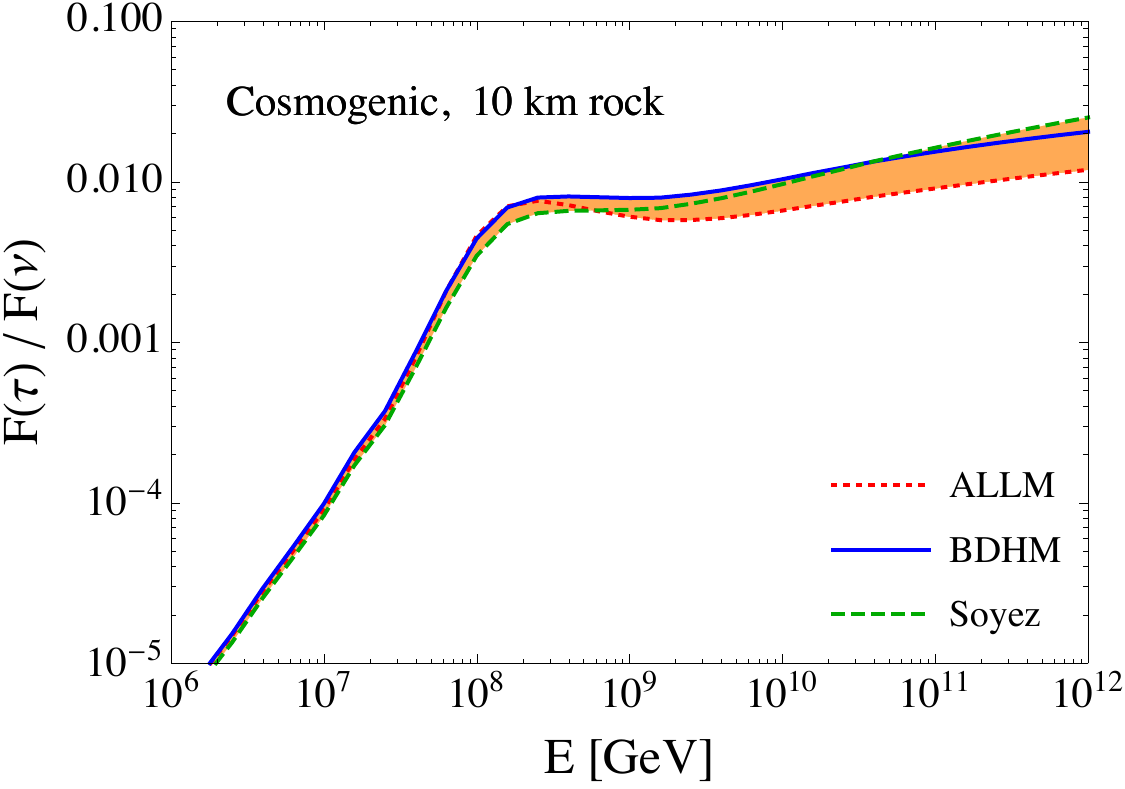}
	\caption{The ratio of the emerging tau flux to an incident  cosmogenic neutrino flux from
	the pure proton cosmic ray model with the AGN evolution scenario of Ref. \cite{Aloisio:2015ega}. }
\label{fig:flux-cosmogenic}
\end{figure}

In Fig. \ref{fig:flux-cosmogenic}, we present the tau fluxes produced from a representative incident cosmogenic neutrino flux. The figure is in the 
same format as Fig. \ref{fig:flux-diffuse-le}. We take the cosmogenic neutrino flux from the pure proton cosmic ray model 
with the AGN evolution scenario shown in  Ref. \cite{Aloisio:2015ega}. 
Other assumptions about the cosmic ray composition and evolution scenarios primarily change the overall normalization of the neutrino flux, so
$F(\tau)/F(\nu)$ is representative for all of the cosmogenic neutrino fluxes in Ref. \cite{Aloisio:2015ega}. 
The tau fluxes from the BDHM and Soyez models have the similar relations to the case of the $E^{-2}$ flux for incident neutrinos.
However, the ALLM transmission function is relatively lower, and it gives rise to the larger uncertainty for the depths of 5 km and 10 km,
with the largest error is still about a factor of 2. In this case, 
the Soyez dipole approach yields a factor of 2.1 larger flux 
than the ALLM approach at the highest energies.

Figs. \ref{fig:flux-diffuse-mc}-\ref{fig:flux-cosmogenic} show the neutrino propagation through short distances relative to the tau range. The tau range goes up to $\sim 50$ km for tau propagation in rock for $E_\tau^i<10^{12}$ GeV, as shown in Fig. \ref{fig:taurange}.
Requiring the emerging tau to have at least 10\% of the incident tau energy reduces the tau range to $\sim 10$ km at $E_\tau^i=10^{12}$ GeV.
The tau lifetime sets another distance scale, namely, $\gamma c\tau = 4.9\ {\rm km}\cdot (E/{10^8\ {\rm GeV}})$. 

For distances less than 10 km in rock, the increase in distance in rock translates to an increase in number of targets for neutrino interactions. At a distance of 10 km of rock, the transmission function for an incident $E_\nu^{-2}$ flux is a factor of between 2.6-3.2 larger than for 2 km of rock for the BDHM parameters. Tau energy loss and decay, depending on the energy, account for the discrepancy between the 
transmission function and a factor of 5 coming from the ratio of the two depths, 2 km and 10 km.
Neutrino attenuation is not significant, even at the highest energies, since the interaction length in rock 
is $\sim 1,500-150$ km for $E_\nu=10^8-10^{12}$ GeV. 

\begin{figure}[htb]
\centering
	\includegraphics[width=0.9 \columnwidth]{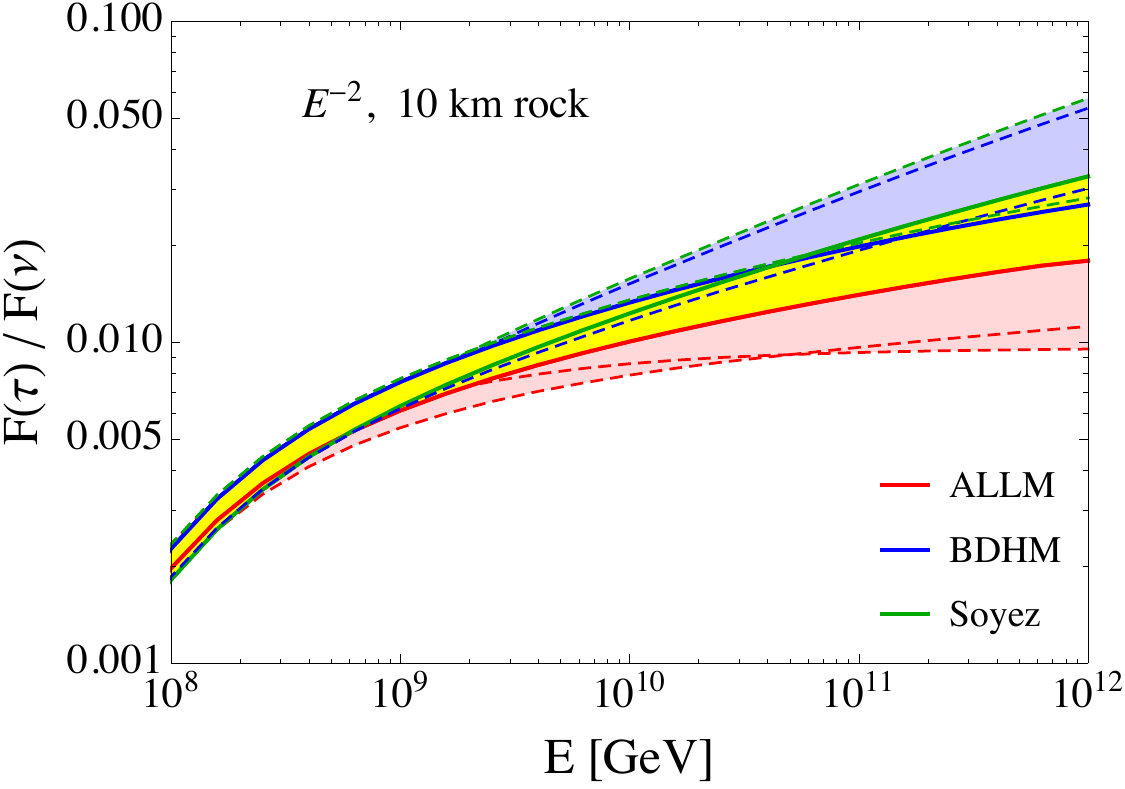}
	\includegraphics[width=0.9 \columnwidth]{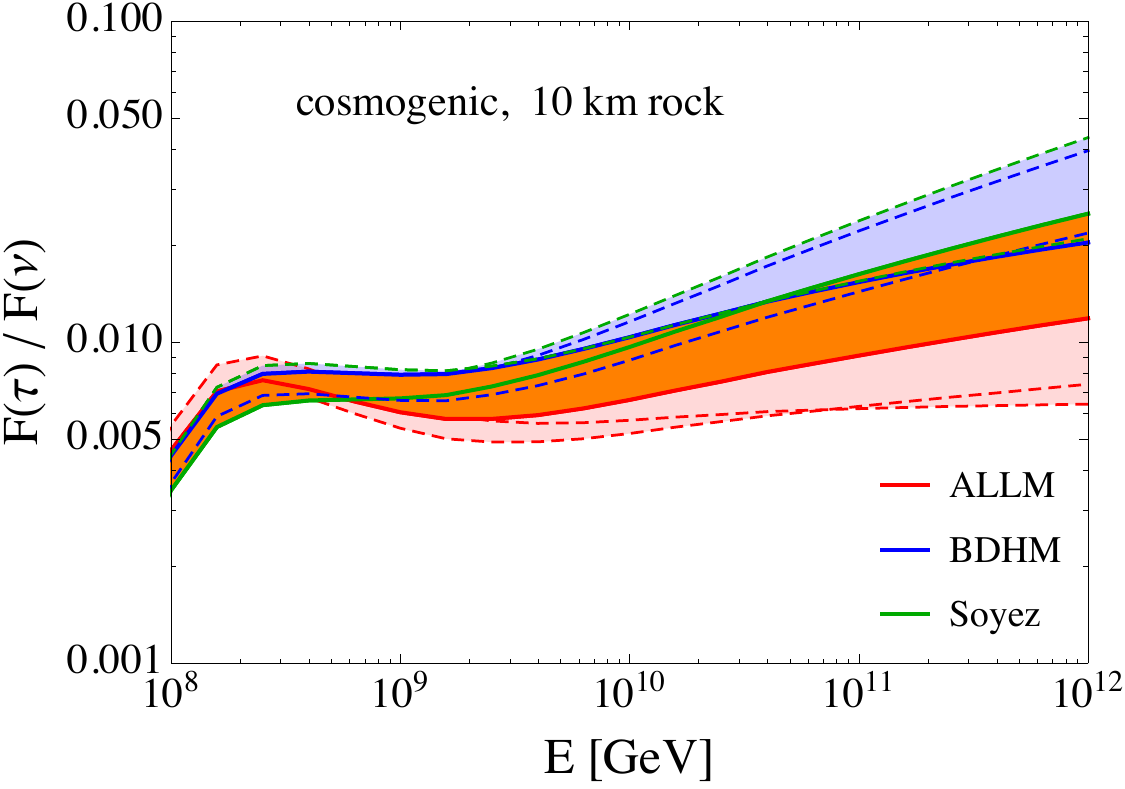}
	\caption{ Uncertainties of the transmission function 
 for 10 km rock
	due to the cross section and the inelasticity for neutrino interaction and the tau energy loss.
	}
\label{fig:errors-10km}
\end{figure}

Fig. \ref{fig:errors-10km} shows the overall uncertainties on the ratio of the emerging tau flux to the incident neutrino flux
from the different approaches for the neutrino cross section, the tau energy loss and the neutrino inelasticity, where a single
approach is not used consistently for both the weak and electromagnetic structure functions.
The solid curves are calculated in consistent approaches, 
hence the yellow and the orange bands present the uncertainties by the approaches as in Fig. \ref{fig:flux-diffuse-le} and Fig. \ref{fig:flux-cosmogenic}.
We also investigated how big is the difference on the results due to the neutrino cross section and the inelasticity from the different approaches using their different combinations. 
In the figures, we present the results that give the largest differences, with the dashed curves for the respective models.
The largest contribution to the uncertainties due to the evaluation in the mixed approaches is from the neutrino cross section. 
The range of neutrino cross sections results in 30\% (50\%) contribution 
towards the 
transmission function uncertainty at 
$E=10^{8}$ GeV ($E=10^{11}$ GeV), 
while the mean neutrino interaction 
inelasticity gives about 5\% and 10\% difference at the same energies.
The overall uncertainties are about 30\% (60\%) at $10^8$ GeV and 
a factor of 3.3 (3.8) at $10^{11}$ GeV for the diffuse $E^{-2}$ flux (cosmogenic flux).

\subsection{Larger distances}

For depths larger than 10 km, one begins to see a saturation in the transmission function, then attenuation effect for even larger depths. In this section, we use the $E_\nu^{-2}$ flux to demonstrate these effects.

\begin{figure}[htb]
\centering
	\includegraphics[width=0.9 \columnwidth]{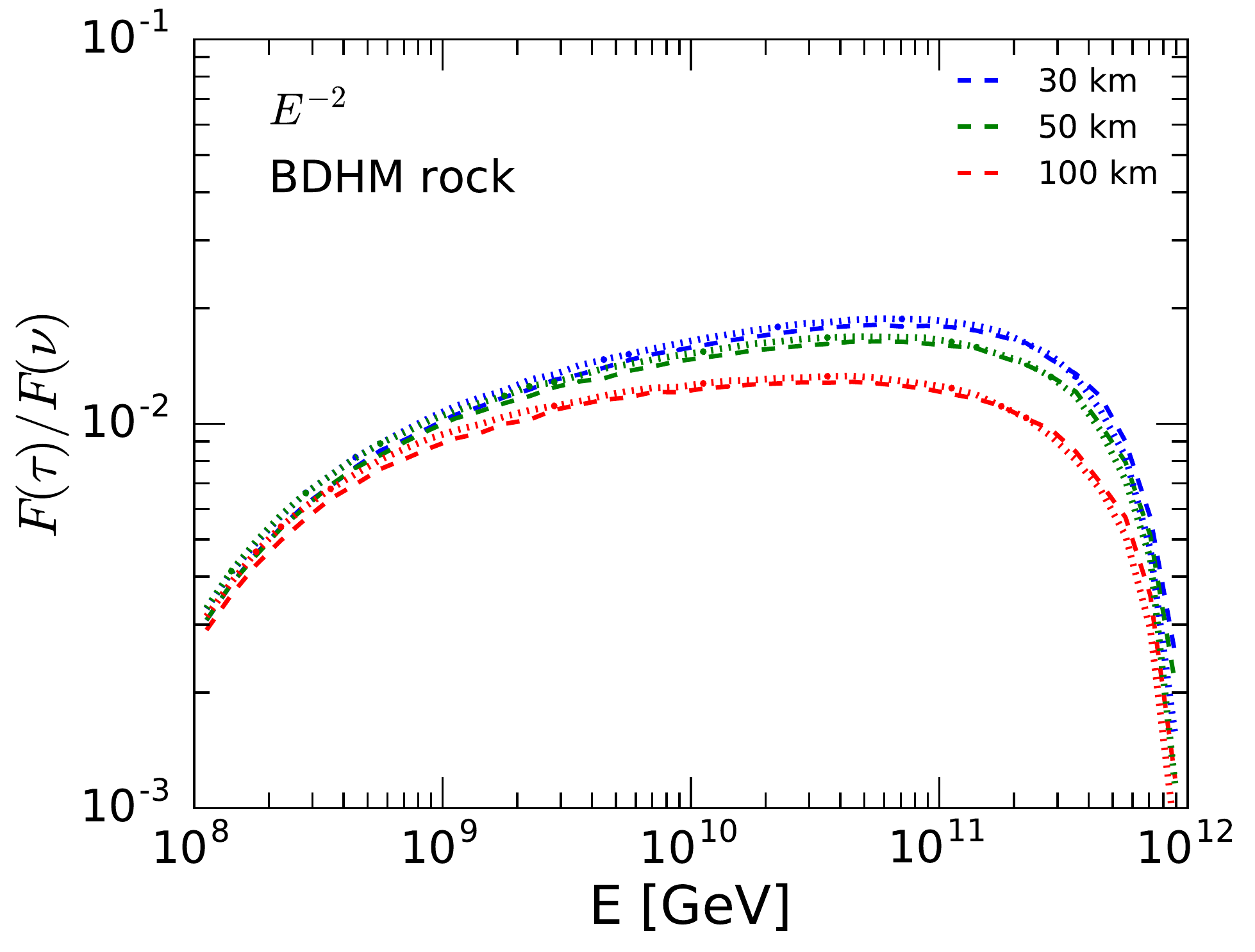}
	\includegraphics[width=0.9 \columnwidth]{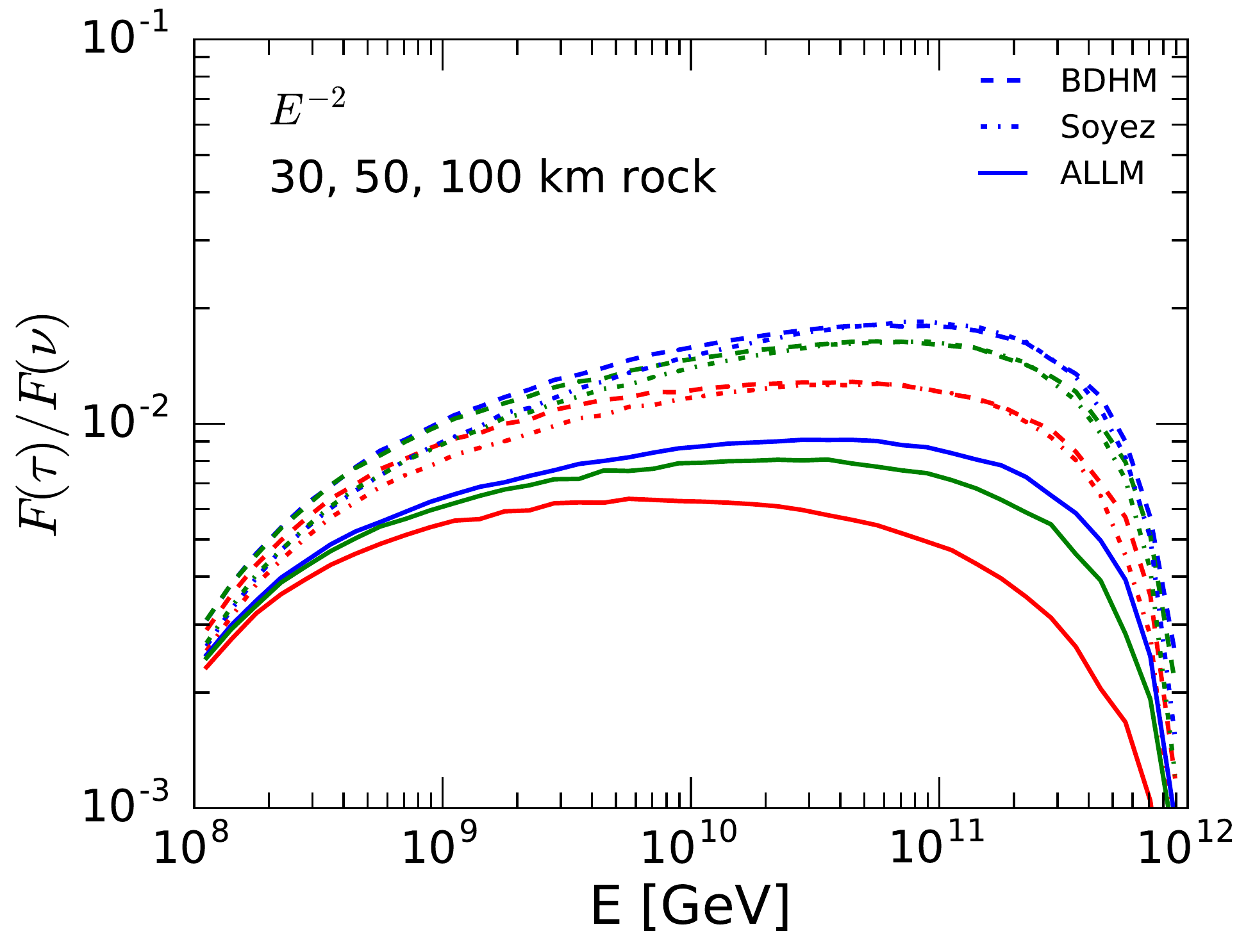}
	\caption{The ratio of the emerging tau flux to an incident neutrino flux scaling as $E_\nu^{-2}$, using the BDHM energy loss and
	neutrino cross section (upper) and all three approaches (lower). In the upper figure, the dashed lines are from the Monte Carlo simulation of neutrino interactions and tau propagation with
	$E_\nu^{\rm max}=10^{12}$ GeV, while the dotted lines use the Monte Carlo program only for the neutrino interaction,
	for 30 (upper), 50 (middle) and 100 (lower curves) km. In the lower figure, we show the BDHM (dashed), Soyez (dot-dashed)  and ALLM (solid) approaches for the same distances. }
\label{fig:flux-em2-larged}
\end{figure}

In the upper plot of Fig. \ref{fig:flux-em2-larged}, we show with the dashed lines the Monte Carlo result for 30 km (upper), 50 km (middle) and 100 km (lower) for $E_\nu^{-2}$ with
$E_\nu^{\rm max} = 10^{12}$ GeV using the BDHM approach.  For an energy $\sim 10^8$ GeV, the transmission function is nearly equal for 30, 50 and 100 km of rock. For this energy, the transmission function is dominated by the decay length $\lambda_\tau(E)$. For much larger distances, the produced tau will decay inside the rock.
The relevant 
target column depth for the incident neutrino is $\sim \rho\lambda_\tau(E)$ which
grows linearly with energy. As the tau energy and decay lengths increase, tau energy loss becomes more important. Then the relevant neutrino target column depth
is $\sim R_\tau(E)\rho$. The transmission function for $30-50$ km at $E\sim 10^9-10^{10}$ GeV is nearly the same, growing with energy because the neutrino cross section increases with energy faster than the energy loss effects degrade the tau energy. For 100 km of rock, 
neutrino attenuation comes in as well, so that the transmission function is  $\sim 10-20\% $ reduced relative to 30-50 km. At higher energies, the curves for 
transmission function begin to further separate
as more neutrinos interact before the final column depth of $\sim R_\tau(E)\rho$. 

The dotted lines in the upper plot of Fig. \ref{fig:flux-em2-larged} show the evaluation of the transmission function where the Monte Carlo program is used only for the neutrino interaction and the maximum neutrino energy is $E_\nu=10^{12}$ GeV. The tau energy $E_\tau^i$ is generated stochastically, smearing the initial tau energy and therefore the maximum $E_\tau$ for the analytic survival probability calculation. For these three distances, the dotted lines match the Monte Carlo results to within 10\%. 
This hybrid method of Monte Carlo simulation for the neutrino interaction and analytic result for the tau propagation
speeds up the numerical comparisons for longer distances, where most of the taus do not emerge after longer distances.

The lower plot in Fig. \ref{fig:flux-em2-larged} shows the transmission function for three different approaches to the energy loss: BDHM, Soyez and ALLM. Increased tau energy loss and some attenuation come into play in the ALLM evaluation compared to the BDHM and Soyez
evaluations.

The Monte Carlo results are in good agreement with the results obtained 
analytically, 
within about 5\% and 10\% at $E=10^{10}$ GeV for the BDHM and Soyez approaches, respectively.  
Above $10^{10}$ GeV, the Monte Carlo results start decreasing due to the cutoff of $E_\nu^{max}=10^{12}$ GeV:
the difference between Monte Carlo and analytic evaluations with no cutoff, at $E=10^{11}$ GeV, 
is about 20\% for BDHM and  about 35\% for Soyez.

\begin{figure}[htb]
\centering
	\includegraphics[width=0.9 \columnwidth]{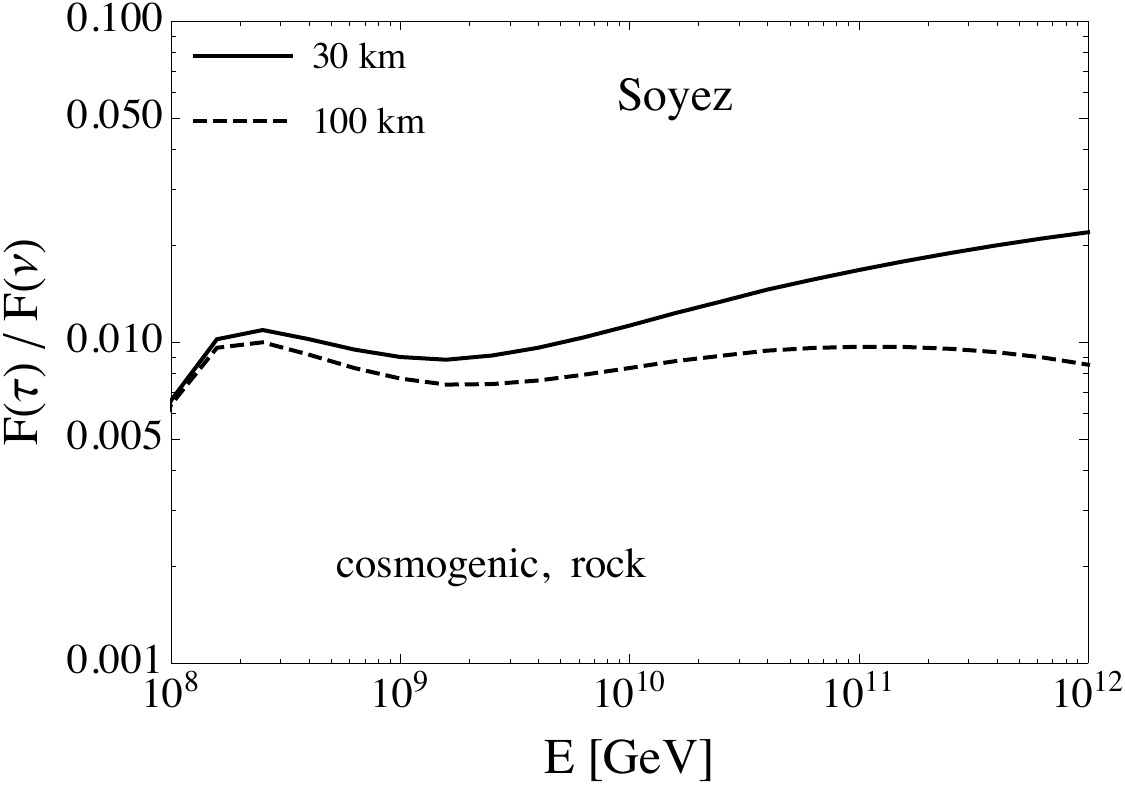}	
	\caption{The transmission function calculated analytically 
 for incident cosmogenic neutrino flux and with tau energy loss in the 
Soyez approach,  for a depth of 30 and 100 km rock.  
	}
\label{fig:flux-larged}
\end{figure}

\begin{figure}[htb]
\centering
	\includegraphics[width=0.9 \columnwidth]{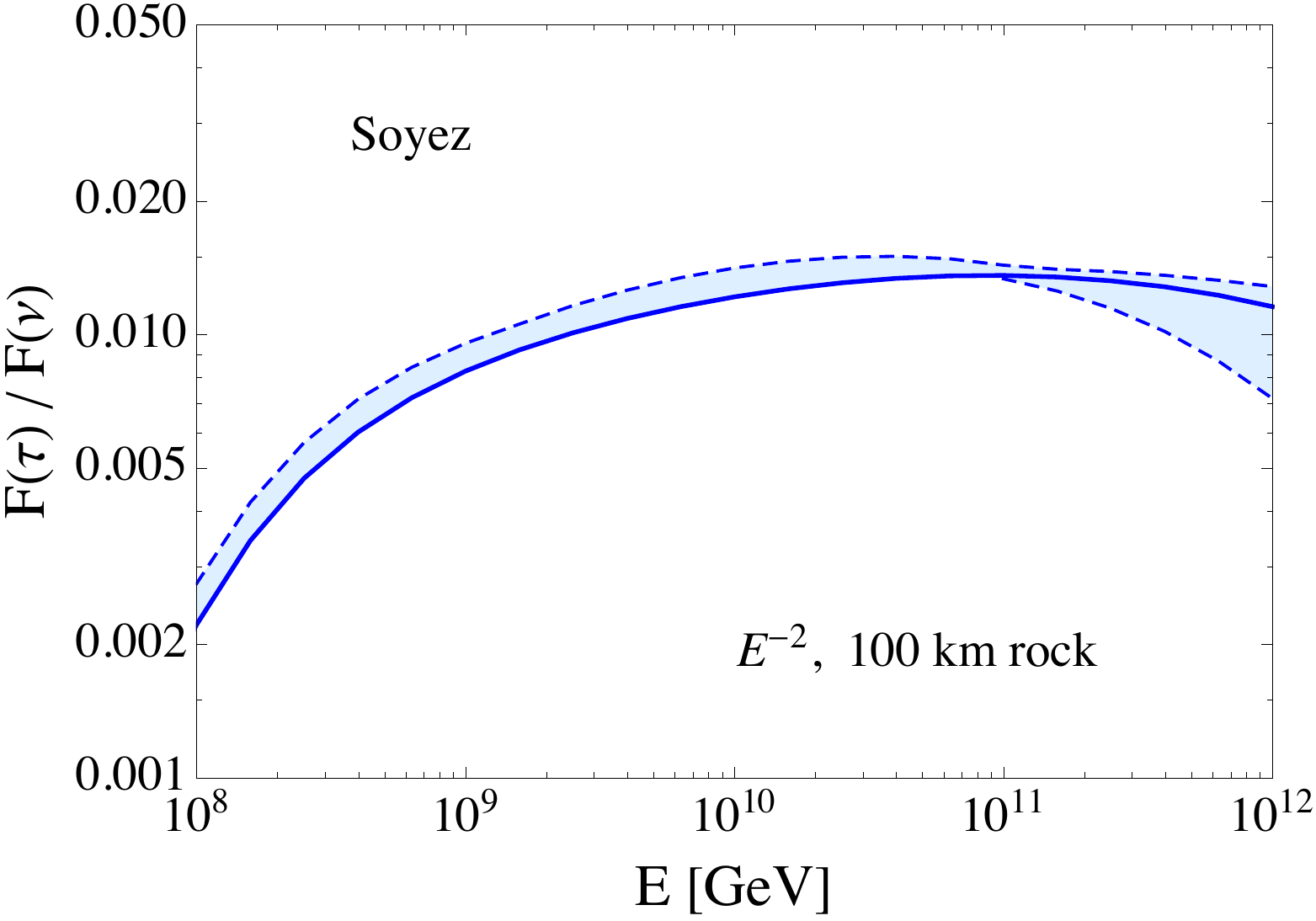}
	\includegraphics[width=0.9 \columnwidth]{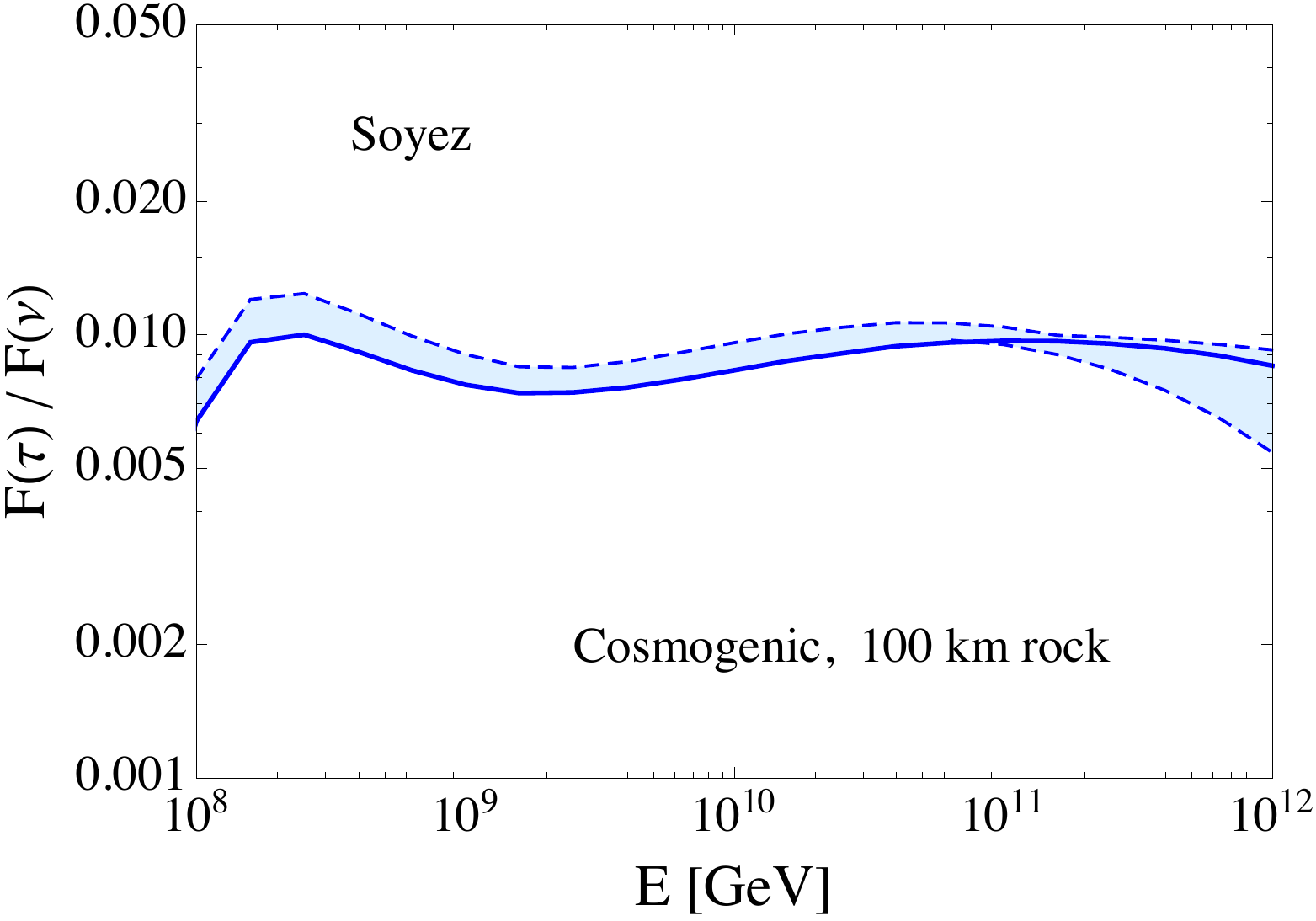}	
	\caption{Uncertainties of the transmission function 
 for 100 km rock 
	due to the neutrino cross section and the neutrino inelasticity for the tau energy loss obtained in Soyez approach.
	}
\label{fig:errors-100km}
\end{figure}

Fig. \ref{fig:flux-larged} shows the transmission function calculated 
analytically for the incident cosmogenic neutrino flux.
For the 100 km, the neutrino attenuation results in the decrease 
of the transmission function at the high energies relative to 30 km. 

Fig. \ref{fig:errors-100km} shows the overall uncertainties on the ratio of the emerging tau flux to an incident neutrino flux
as in Fig. \ref{fig:errors-10km}, but for the larger distance of 100 km.  The 
band shows the impact of including the range of neutrino cross sections and 
neutrino inelasticity 
obtained in all three approaches for the tau energy loss obtained with the 
Soyez approach.  
The overall uncertainties in this figure, for the energy range of 
$10^8 $ GeV$\leq E \leq 10^{11}$ GeV, are within 35\% 
for both $E_\nu^{-2}$ and cosmogenic neutrino fluxes.
The largest difference is at energy of about $10^{10}$ GeV, about 32\% and 34\% 
for $E_\nu^{-2}$ and cosmogenic neutrino fluxes.  
At  $E = 10^{11}$ GeV, the uncertainty reduces to 24\% and 30\%, respectively.
From Fig. \ref{fig:flux-em2-larged} we note that for $E_\nu^{-2}$ neutrino 
flux and distance of 100 km rock, the different approaches to tau energy 
loss give larger 
uncertainty than the uncertainty due to 
neutrino cross section or inelasticity. For $E=10^8$ GeV, 
the transmission functions differ by a factor of 1.3, while
for $E=10^{10}$ GeV, the BDHM gives a factor of 2.0 larger result 
that the one evaluated in ALLM approach.

\begin{figure}[htb]
\label{fig:chordlength}
\centering
\resizebox{0.85\columnwidth}{!}{
\begin{tikzpicture}
 [
   scale=1.,
  point/.style = {draw, circle, fill=black, inner sep=0.1pt},
 ]
\def\rad{5cm}
\node (C) at (0,0) [point]{};
\draw[line width=1.5pt] (C) circle (\rad);
\node (P)  at +(120:\rad)  [point]{};
\node (R)  at +(93:\rad)  [point]{};
\node (R2)  at +(174:\rad)  [point]{};
\draw[->,line width=1.5pt] ($(P)!1.4!-68:(C)$)--($(P)!0.9!90:(C)$) ;
\draw(-1.9,3.45 ) node{\Crossx} ;
\draw(-2.2,2.5) node{\Large $D$}
(-3.6,2.4) node{\Large $\nu_\tau$}
(-1.3,4.4) node{\Large $\tau$};  
\draw[line width=1.pt, dashed, black] (C) --(R)-- ([turn]-0:2cm) 
(0,2.4) node[right]{\Large $R_\oplus$};
\draw[line width=1.pt, dashed, black] (C) --(R2);
    \draw[-,line width=1pt](-0.28,5.7)arc(100:35:.50) 
    (0.1,5.7) node[above]{\Large $\theta$};
    \draw[-,line width=1pt](0.36,5.57 )arc(80:-30:.40) 
    (1,5.2) node[above]{\Large $\alpha$};    
\end{tikzpicture}
}
\caption{Chord length through the Earth $D$ for nadir angle $\theta$ and radius of the Earth $R_\oplus$.}
\end{figure}
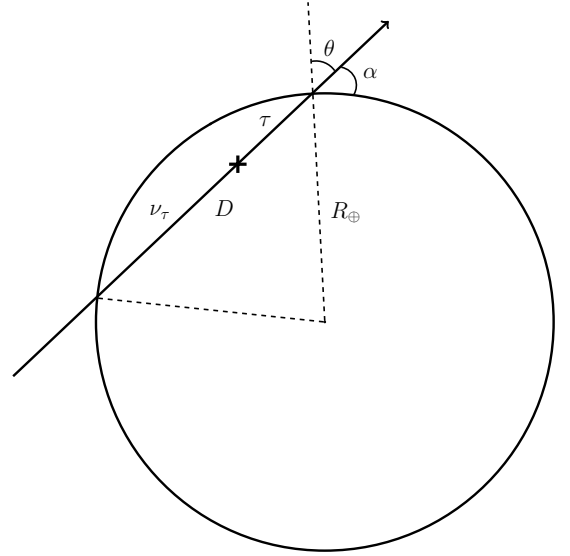

A distance of 100 km of rock corresponds to an angle of approximately $0.5^\circ$ below the horizon.
Fig. 16 shows schematically the chord length $D$ as a function of nadir angle $\theta$, where
\begin{equation}
D = 2 R_\oplus \cos\theta\ .
\end{equation}
For a nadir angle of $85^\circ$, the chord length is more than 1,000 km. For neutrinos with energies larger than a few times $10^8$ GeV, the neutrino interaction length is smaller than the chord length for nadir angles much less than $85^\circ$, so the transmission function will be appreciable only for nadir angles close to $90^\circ$. We define the angle $\alpha$ relative to the horizon:
\begin{equation}
\alpha = \frac{\pi}{2}-\theta\ .
\end{equation}
For reference, we show the chord lengths and maximum depth of the neutrino/tau trajectory below the Earth's surface (sagitta) in Table \ref{table:chord}.

\begin{table}[h]
\caption{Path length and sagitta to correspond to the angle $\alpha$ relative to the horizon.}
\begin{center}
\begin{tabular}{|c|c|c|}
\hline
$\alpha$  [$^\circ$]& Sagitta [km]& Path length [km]\\
\hline 
\hline 
1 & 0.97 & 222 \\
\hline 
2 & 3.88 & 445  \\
\hline 
3 & 8.73 & 667\\
\hline 
4 & 15.5 & 889\\
\hline 
5 & 24.2 & 1,110 \\
\hline 
6 & 34.9 & 1,330 \\
\hline
\end{tabular}
\end{center}
\label{table:chord}
\end{table}%

\begin{figure}[htb]
\centering
	\includegraphics[width=0.9 \columnwidth]{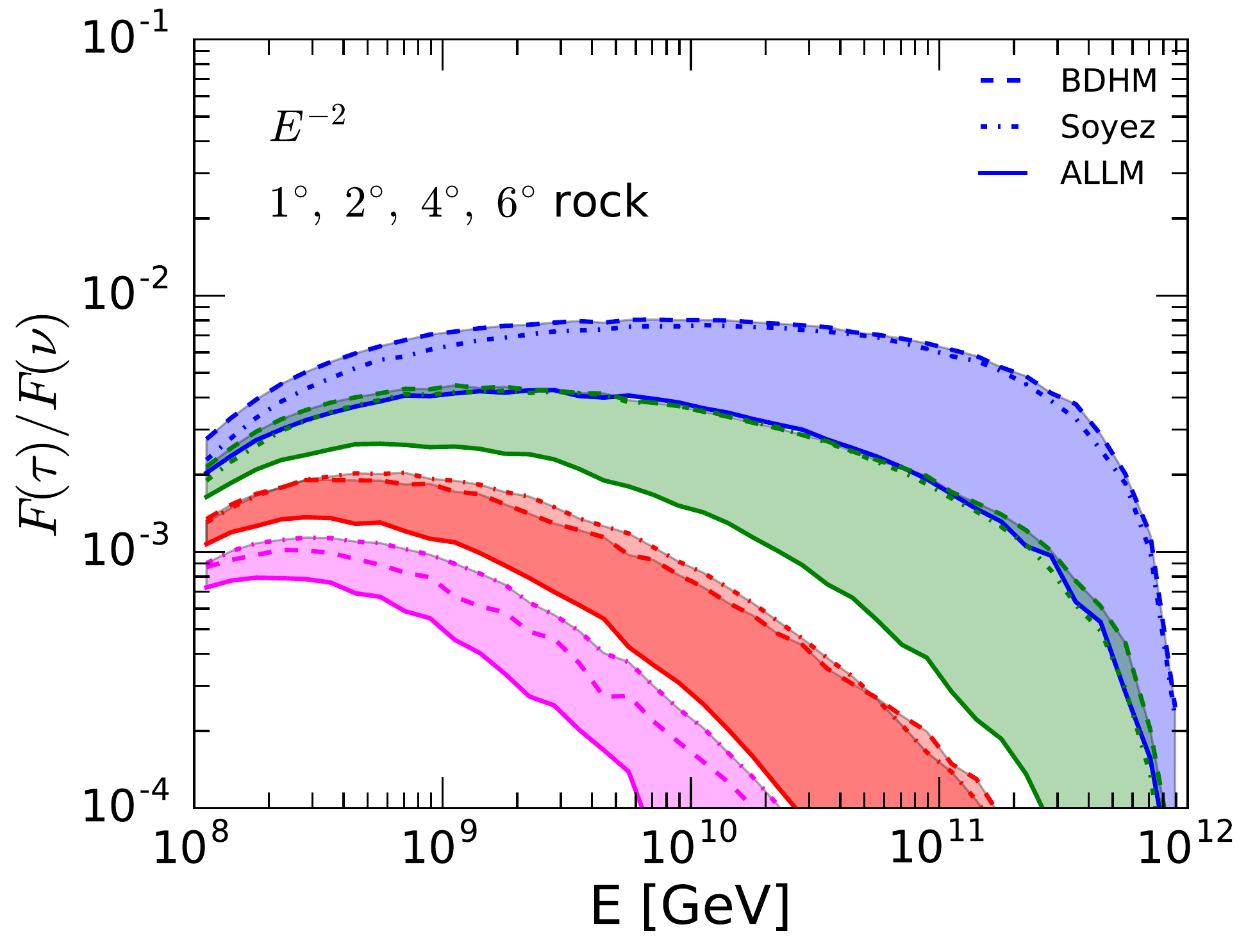}
	\caption{The transmission function for an incident neutrino flux scaling as $E_\nu^{-2}$ for $E_\nu<
	10^{12}$ GeV for $\alpha = \pi/2-\theta = 1^\circ, \ 2^\circ,\ 4^\circ$ and $6^\circ$ for the BDHM, Soyez and ALLM approaches.}
\label{fig:angle-transmission}
\end{figure}
 In Fig. \ref{fig:angle-transmission}, we show a plot of $F(\tau)/F(\nu)$ for different angles $\alpha$ relative to the horizontal using
 $R_\oplus=6.37\times 10^3$ km. The surface rock-mantle interface is several
 hundred kilometers below the Earth's surface \cite{PalomaresRuiz:2005xw}, so for $\alpha \leq 6^\circ$, $\rho=2.65$ g/cm$^3$ is a good approximation for a detector on land.
Each curve represents the transmission function for a given approach and the shaded band shows the spread in predictions for $F(\tau)/F(\nu)$
for tau neutrinos skimming with $\alpha = 1^\circ, \ 2^\circ ,\ 4^\circ$ and $6^\circ$ below the horizon.
Similar to the shorter trajectories, the span of predictions at $E=10^8$ GeV is less than at $E=10^{10}$ GeV and higher. At $E=10^8$ GeV,
$F(\tau)/F(\nu)$ from the BDHM approach is a factor of $1.2-1.4$ higher than for the ALLM calculation for all of the angles shown.
By $E=10^{10}$ GeV, the BDHM transmission function is a factor of $2.2 \ (1^\circ)$ to $13\ (6^\circ)$ higher.

How the uncertainty in the transmission function translates to a particular measurement depends on the relative importance of different
energy regions and angles. If one considers air showers that must develop before an altitude $h=20$ km and taking the time dilated decay length as the distance the tau travels after emerging from the Earth at an angle $\alpha=1^\circ-6^\circ$, then $E_\tau<8.3\times 10^9$ GeV. For neutrino fluxes
that decrease with energy, there is a further suppression of the high energy regime.

\begin{figure}[htb]
\centering
	\includegraphics[width=0.9 \columnwidth]{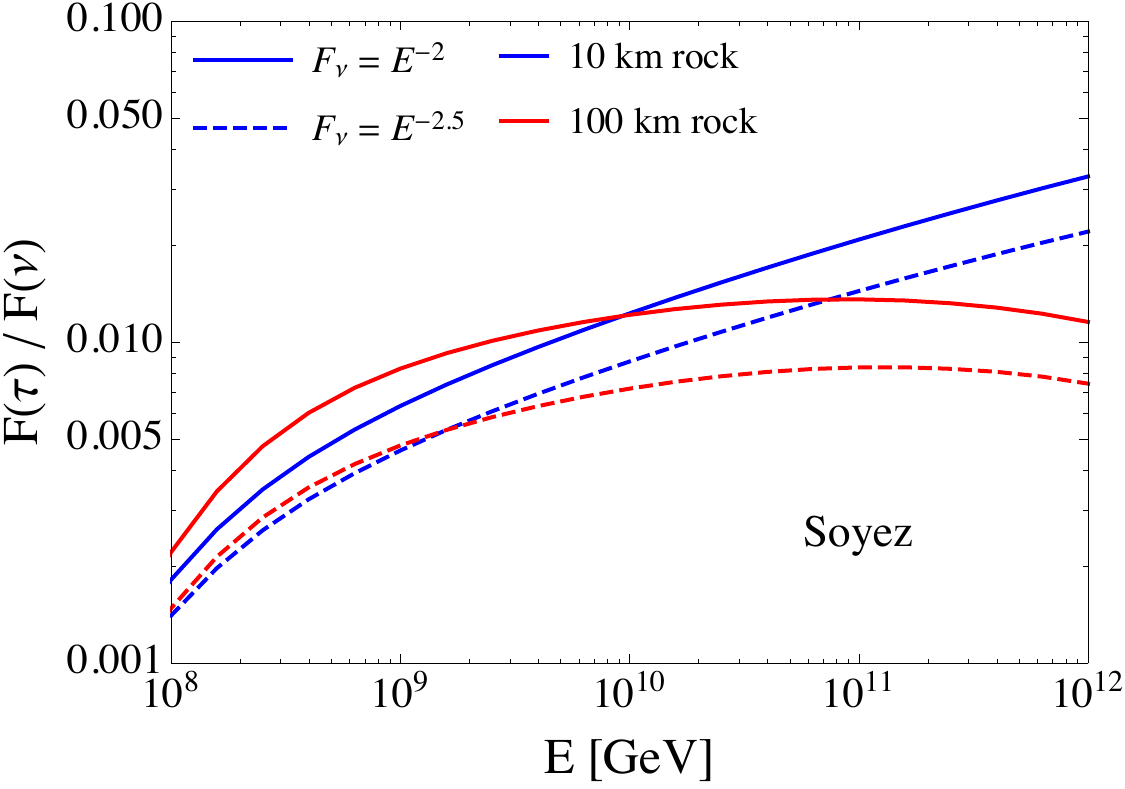}	
	\caption{Comparison of the transmission functions for 
incident neutrino flux scaling as $E_\nu^{-2.5}$ and $E_\nu^{-2}$ 
	 for 10 km and 100 km.
	}
\label{fig:flux-e2p5}
\end{figure}

Before we turn to the specific example of the Pierre
Auger Observatory, we discuss the impact of different spectral indices on the range of predictions for the transmission function.
In Fig. \ref{fig:flux-e2p5}, we present the transmission function for the incident neutrino fluxes 
scaling as $E_\nu^{-2.5}$ and $E_\nu^{-2}$  taking the Soyez approach for the reference. 
The transmission function for the steeper neutrino flux is smaller.  For example, the $E_\nu^{-2.5}$ incident neutrino flux gives transmission function 
 about 23\% (33\%) lower than in case of incident neutrino flux being 
$E_\nu^{-2}$ at $E=10^8$ GeV for the 10 km (100 km) distance.
This reduction effect becomes larger as energy increases, 
 at $E=10^{11}$ GeV it is about 30\% (40\%).  
We find effect to be the same for the BDHM approach, 
while for the ALLM approach it is 27\% (45\%) at the corresponding energies for the 10 km distance.

\section{Application: Pierre Auger Observatory}

Our results are directly applicable to an approximate evaluation of the effective aperture of the Pierre Auger Observatory. Following Refs. \cite{Aramo:2004pr,Miele:2005bt,Abraham:2009uy}, we use the analytic survival probability for the 
kernel $K(E_\nu,D,E_\tau)$, involving an integration of Eq. (\ref{eq:ftau-results}) to get
\begin{equation}
F_\tau(E_\tau, D) = \int dE_\nu F_\nu(E_\nu,0) \, K(E_\nu, D, E_\tau)\ .
\end{equation}
To simplify the expression, we use $\langle y\rangle = y_0=0.15$ for all energies. The $\delta$-functions relating initial and final tau energies and the neutrino energy to the initial tau energy simplify the expression of the kernel. The kernel for the 2-parameter fit of $\beta$ appears in Appendix B.

\begin{figure}[htb]
\centering
\resizebox{0.9\columnwidth}{!}{
\begin{tikzpicture} 
\path   
    (5,0) coordinate (A);
 \fill[gray!15] (A) arc (60:100.5:18);
 \draw[black,line width=1pt] (A) arc (60:100.5:18);
 \fill[gray!15,rounded corners] (-7.25,2.1)--(-7.25,-0)--(A);
  \draw[purple,line width=1.5pt](-7.25,1.5)--(0.2,2.91) 
    (0.24,2.92)node{\Cross} (0.2,3) node[above,black]{\Large $\tau$ decay};
  \draw[purple, dashed,line width=1.5pt](0.2,2.91)--(3.5,3.58)
     (2.,3.2) node[below,black]{\Large $x_0$};
  \draw[black,line width=0.8pt](3.4,3.55)--(2.5,1.19) (3.5,2.2) node[left]{\Large $h$};
    \draw[black,line width=0.5pt](-2.57,4)--(-2.7,2.366);
    \draw[black,line width=0.5pt](-5,2.55)--(-0.2,2.166);    
    \draw[-,line width=1pt](-2.67,2.75)arc(90:13:.38) 
      (-2.,2.95) node[left]{\Large $\theta$};
\end{tikzpicture}
}
\caption{The geometry used to approximate the Auger aperture. Here, $h$ is the altitude a distance $x_0$ from the tau decay point, for taus emerging from the Earth with a nadir angle $\theta$.}
\label{fig:Auger-geometry}
\end{figure}
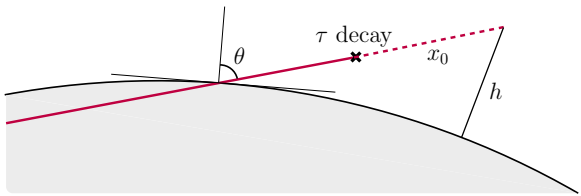

The aperture depends on the tau energy and lifetime. Ref. \cite{Abraham:2009uy} suggests that the relevant altitude
for detection is about $x_0=10$ km along the shower axis from the point of the tau decay. 
The kernel is translated to the effective aperture (approximately) by first evaluating the differential probability to 
have a tau decay a distance $x-x_0$ from exiting from the Earth, as a function of altitude $h$ above the Earth's surface at
$x$. The configuration is shown schematically in Fig. \ref{fig:Auger-geometry}.

The distance $x$ and altitude $h$ are related to the nadir angle $\theta$, as is the  chord length $D$ that first the
tau neutrino, then the tau, travels through the Earth. We take a constant Earth density $\rho=2.65$ g/cm$^3$
 for the angles of relevance,
namely $\theta=\pi/2-\alpha\to \pi/2$ for $\alpha=0.1$ rad. The distances $x$ and $h$, and nadir angle $\theta$, are related by
\begin{equation}
(R_\oplus+h)^2 = R_\oplus^2+x^2+2 R_\oplus x\cos\theta \ .
\end{equation}
The probability density is
\begin{eqnarray*}
\frac{d^2 P_\tau(E_\nu,E_\tau,\cos\theta,h)}{dE_\tau \, dh} 
&=& K(E_\tau,\cos\theta,E_\nu)\frac{e^{-\frac{(x-x_0)}{\lambda_{dec}}}}{\lambda_{decay}}\\
&\times & \frac{h+R_\oplus}{\sqrt{R_\oplus^2\cos^2\theta + h^2+2R_\oplus h}}
\end{eqnarray*}
where $\lambda_{dec} = E_\tau c\tau_\tau/m_\tau$ is the usual decay length. 

We evaluate the effective aperture by integrating over the fraction of the solid angle that is important, over the tau energies
and over altitudes weighted by an approximate efficiency for detection, a product of the identification and trigger efficiencies. 
The maximum efficiency for identification and trigger is each at 0.84 since the muonic decay of the tau does not produce a shower. In Ref. \cite{Abraham:2009uy}, efficiencies for selected energies are shown and described as being primarily functions of altitude and tau energy. We approximate the effective efficiency by an energy dependent step function for $h$,
\begin{eqnarray*}
\epsilon_{eff}(E_\tau, h)&\simeq & 0.64\ \theta(h_{max}(E_\tau)-h)\\
h_{max} &=& 0.5\log_{10}(E_\tau/10^8\ {\rm GeV}) \ .
\end{eqnarray*}
This gives the effective aperture in the analytic approximation,
\begin{eqnarray*}
Ap(E_\nu) &=& 2\pi A  \int \sin\theta\cos\theta\, d\theta\, \int dE_\tau\, \int dh \, \frac{d^2 P_\tau}{dE_\tau \, dh} \\
&\times & \epsilon_{eff}(E_\tau, h)\ .
\end{eqnarray*}

\begin{figure}[htb]
\centering
	\includegraphics[width=\columnwidth]{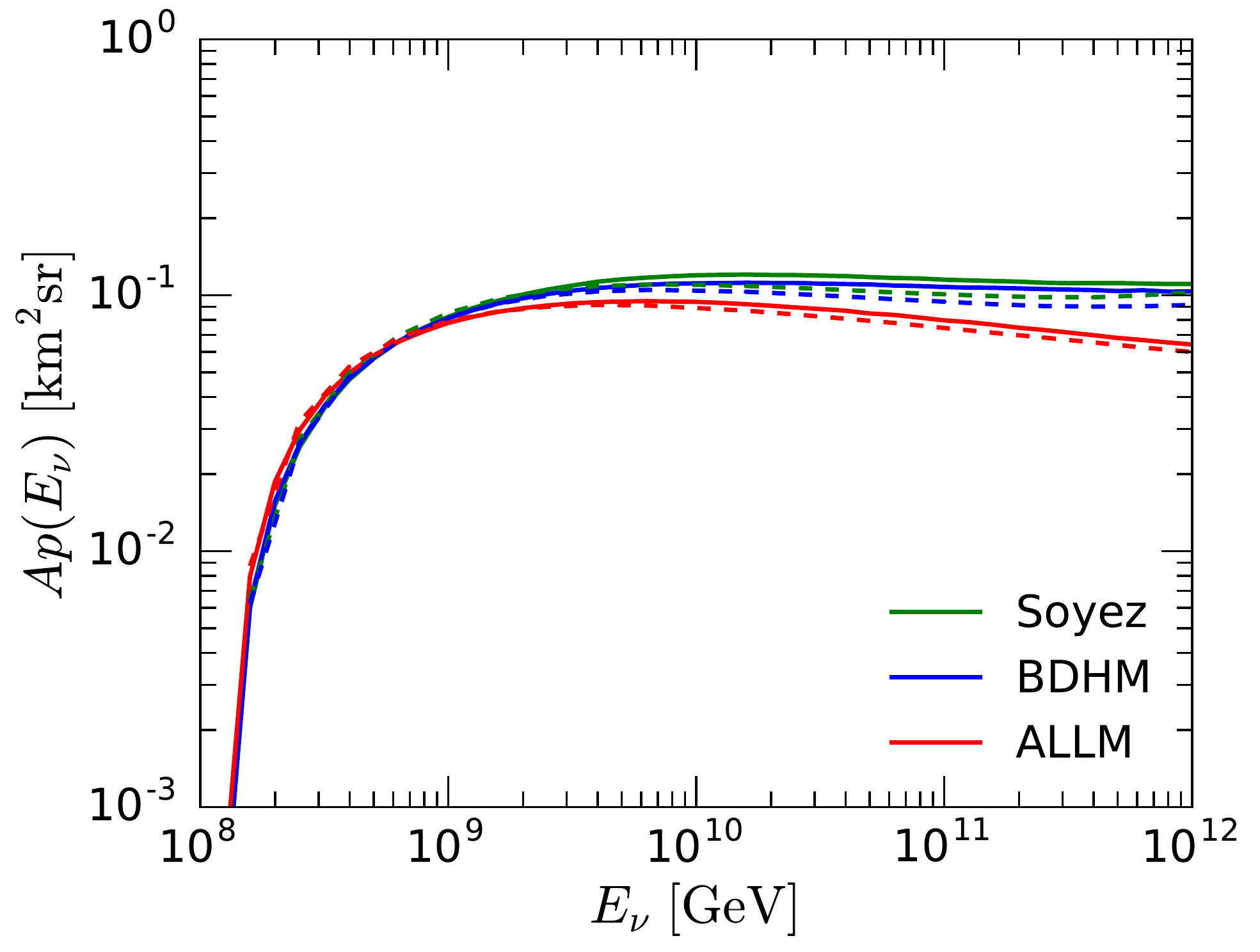}
	
	\caption{The effective aperture for  the  Pierre Auger Observatory
	for $A=3,000$ km$^2$ using the analytic approximation to the tau survival probability in rock and an approximate efficiency for detection, for Soyez, BDHM and ALLM models for both energy loss and neutrino cross section (solid line), and with the NLO perturbative neutrino cross section substituted (dashed lines).}
	\label{fig:aperture}
\end{figure}

The evaluation of the effective apertures for the Pierre Auger Observatory  with $A=3,000$ km$^2$ are shown with the solid lines for Soyez, BDHM and ALLM approaches from top to bottom in Fig. \ref{fig:aperture}. Our results here, with the crude approximation for the efficiency, are within a factor of $\sim 4$ of those of Ref. \cite{Miele:2005bt}, where a careful accounting of the local 
topography near the Observatory was included. Our approximations serve to test the impact of the different approaches
to energy loss and neutrino cross sections.
In the range of $E_\nu=10^9-10^{12}$ GeV, the Soyez and BDHM evaluations differ by 
between $2-7\%$. At $E=10^9$ GeV, the ALLM curve is about $6\%$ lower, however ALLM aperture drops with higher energies, so that the Soyez aperture is a factor of 1.7 times larger than the ALLM aperture.

The dashed lines in the figure show the impact of including a parameterized  NLO perturbative neutrino nucleon
cross section from Ref. \cite{Jeong:2010za}.
For $E_\nu=10^9-10^{12}$ GeV, the ratios of Soyez to ALLM apertures are from 1.04 to 2.0.
For the event rates for an $E^{-2}$ incident tau neutrino flux, the peak of the differential event rate is at $E_\tau$ a few times $10^8$ GeV, where differences in the approaches are small. Our approximate calculations have the total event rates differing by only a few percent.

\section{Conclusions}

We have calculated the emerging tau flux generated from tau 
neutrino interactions followed by the tau propagation through rock.  
We have evaluated neutrino charged current cross section and the 
tau energy loss in 
various QCD-motivated approaches, such as dipole approach as 
well as the more phenomenological approach that determines 
neutrino cross section using the same $F_2$ parameterization 
used in evaluating
the tau energy loss parameter $\beta_\tau^{\rm nuc}$.  Our comparisons of
results obtained analytically with those obtained with the 
Monte Carlo treatment of the tau energy loss show good agreement for fluxes that
scale like $E_\nu^{-2}$. While not a fundamental result, the analytic treatment of
tau propagation in materials is a useful tool to survey different approaches to calculate
tau electromagnetic energy loss.
We find that the tau electromagnetic energy loss at PeV range can 
be well described by $\beta_{\rm fit}(E) = 
\beta_0+\beta_1 \ln (E/E_0)+\beta_2 \ln^2 (E/E_0)$ 
with the parameters in Tables \ref{table:betatau} and \ref{table:beta-3par}. For the BDHM and
dipole model treatment of electromagnetic energy loss, $\beta_2\simeq 0$ is a good approximation.

We have also parameterized our results for the neutrino 
charged current cross section for each approach we considered, 
reproducing the numerical result to within $\pm 5\%$.  
Finally comparisons of the $\nu_\tau$ fluxes after propagation over 
various distances in rock, between Monte Carlo simulations and 
analytic calculations for each theoretical approach, are in 
good agreement in the energy regime where the Monte Carlo upper limit on the neutrino energy has no impact.  The parameterization of the survival probability and energy relations are useful
tools to determine the implications of new models or approaches without a full
Monte Carlo simulation. 

The different approaches to the tau energy loss calculation have larger impacts at high energies
and longer trajectories in the Earth.  For example, different approaches to electromagnetic energy loss can give  transmission 
functions that differ by a factor of $\sim 10$ for $D\sim 1,000$ km.  Nevertheless,
our results for the Pierre Auger Observatory aperture, with a simplified efficiency, show that event rates with an
$E_\nu^{-2}$ incident neutrino flux are minimally uncertain due to the energy loss and the neutrino cross section, because
at $E_\tau\simeq 10^9$ GeV, the approaches considered here give very similar results.

In addition to applying our results to evaluating the 
effecting aperture for the 
Pierre Auger Observatory, our results can be applied to 
 the High Altitude Water Cherenkov (HAWC) observatory and the 
All-sky Survey High Resolution Air-shower detector (ASHRA).  
For HAWC, the propagation distance is 0.35 to 6 km (width of the 
volcano at summit and at the observatory elevation) \cite{Vargas:2016hcp}.  
We have compared results using the Monte Carlo simulation to analytic 
calculation of the tau flux exiting the Earth performed using the approaches
presented in this paper for the above distances. 
Additionally, one can also consider the 
effective aperture for HAWC, given the 
approximate distance of 4 km from the edge of the volcano to 
the HAWC array and the array surface area of $22,000$ m$^2$. Since HAWC is sensitive to showers that pass through detector elements coming from neutrino conversions in 
the volcano, the effective solid angle is much smaller than for the Pierre Auger Observatory.  Similar 
conditions can be found in the Ashra configuration proposed to be 30 km from Mauna
Kea, with a propagation distance of approximately 20 to 25 km.  
One can adopt our Auger aperture calculation with $\beta_\tau$ 
parameterizations along with the analytic computations for the 
tau neutrino propagation to predict the induced events at the detector arrays.
The calculational technique discussed here could also be applied to a space-based telescope pointing towards the Earth as discussed in, e.g., Ref. \cite{Neronov:2016zou}.

By using a consistent low-$x$ treatment of both energy loss and the cross section, we have shown that the
different parameterizations of $F_2$ and dipole approach yield similar emerging tau fluxes for short distances and for
tau energies $\sim 10^8-10^9$ GeV. By using a separate
NLO parton model calculation of the neutrino cross section, the range of emerging tau flux predictions is broader for the 
approaches considered here. Overall, for large distances, the differences in modeling the tau energy loss account for most of the 
differences in the calculated transmission functions. As forward, low invariant mass processes are studied in more detail, small-$x$
parton distribution functions will help constrain $F_2$ at $Q^2$ values where the parton model applies. The ALLM
parameterization gives a larger tau energy loss at high energies than the dipole model or BDHM parameterization.
If the ALLM parameterization or the BDHM parameterization were ruled out experimentally, the theoretical uncertainty associated with ultrahigh
energy skimming tau neutrinos would be reduced. 

\begin{acknowledgments}
We thank J. Alvarez-Muniz and G. Parente for discussions.
This research was supported in part by the US Department of Energy contracts DE-SC-0010113,  DE-FG02-13ER-41976 and DE-SC0009913.
\end{acknowledgments}

\begin{appendix}

\section{Dipole model}

In the dipole model, the electromagnetic structure function $F_2(x,Q^2)$ is written in terms of the photon wave functions 
and the dipole cross section \cite{Nikolaev:1990ja} 
\begin{eqnarray}
\nonumber
F_2(x,Q^2) &=& \sum_f  \frac{Q^2}{4\pi^2 \alpha_{em}}  \int d^2r \int_0^1 dz \Biggl[ \mid \Psi_{L}^{(f)}(r,z;Q^2) \mid^2\\
&+&  \mid \Psi_{T}^{(f)}(r,z;Q^2) \mid^2 \Biggr] \sigma_{\rm dip}(r,x) \ . 
\end{eqnarray}
The photon wave functions, in terms of the transverse separation $r$ and the fractional momentum $z$, are given by  
\begin{eqnarray}
\label{eq:psil}
|\Psi_L^{(f)}(r,z ;Q^2) |^2&=&e_f^2\frac{\alpha_{em} N_c}{2\pi^2} 4Q^2 z^2(1-z)^2K_0^2(r \bar{Q}_f)\\
\nonumber
|\Psi_T^{(f)}(r,z;Q^2) |^2&=&e_f^2\frac{\alpha_{em} N_c}{2\pi^2}\Bigl( [z^2+(1-z)^2]
\bar{Q}_f^2 K_1^2(r \bar{Q}_f)\\
&+& m_f^2 K_0^2(r \bar{Q}_f)\Bigr) 
\label{eq:psit}
\end{eqnarray}
using the modified Bessel functions $K_0$ and $K_1$, the quark electric charge $e_f$ and mass $m_f$.
Here,  $\bar{Q}_f^2=z (1-z) Q^2+m_f^2$, and we use $m_u=m_d=m_s=0.14$ GeV, $m_c=1.4$ GeV and $m_b=4.5$ GeV. 
The electromagnetic fine-structure constant is labeled $\alpha_{em}$ and
$N_c=3$ is the number of colors in Eqs. (\ref{eq:psil}-\ref{eq:psit}).
For the dipole cross section, $\sigma_{\rm dip}(r,x)$, we use the parameterization in Ref. \cite{Soyez:2007kg} by Soyez.

\section{Three parameter fit to $\beta_\tau(E)$}

In order to extend our investigation to lower energies, a better parameterizations for the energy loss $\beta_\tau$
than Eq. (\ref{eq:beta2})
is useful. Adding a $\ln^2 E$ term, we use
\begin{equation}
\label{eq:beta-3par}
\beta_\tau ^{\rm nuc} (E) = \beta_0 + \beta_1\ln(E/E_0) + \beta_2\ln^2(E/E_0) 
\end{equation}
with $E_0=10^{10}$ GeV. Results for the parameters are in Table \ref{table:beta-3par}.
The results for the BDHM and Soyez approaches
are matched well with the output from the direct calculation. 
The differences are mostly less than 1\% between $ E= 10^6  - 10^{12}$ GeV.
The ALLM fit is not as good as the other models, especially at low energies, 
but it works well, to within 5 \% between  $E = 2  \times 10^7  - 5\times 10^8$ GeV.
Between $E = 5  \times 10^8  - 10^{11}$ GeV, it works better, to  within 3 \%.

\begin{table}[h]
\caption{$\beta_\tau^{\rm nuc}$ in [$10^{-6}$ cm$^2$/g] using Eq. (\ref{eq:beta-3par}) for
$E=10^6-10^{12}$ GeV.}
\begin{center}
\begin{tabular}{|c|c|c|c|}
\hline
Model & $\beta_0^{\rm nuc}$ & $\beta_1$ & $\beta_2$\\
\hline 
\hline 
BDHM & 0.425 & 4.04 $\times 10^{-2}$ & 1.12 $\times 10^{-3}$\\
\hline 
Soyez& 0.371 & 3.20 $\times 10^{-2}$ &9.54 $\times 10^{-4}$\\
\hline 
Soyez-ASW & 0.461 & 3.90 $\times 10^{-2}$ & 1.13 $\times 10^{-3}$\\
\hline 
ALLM & 1.02 & 0.210 & 1.51$\times 10^{-2}$\\
\hline
\end{tabular}
\end{center}
\label{table:beta-3par}
\end{table}%

Starting from the energy dependence of $\beta_\tau$, the relation between the
initial tau energy $E_\tau^i$ and distance traveled $z'$ to the final tau energy $E_\tau$ is
\begin{eqnarray}
\label{eq:etau-3par}
E_\tau &=& \exp \Biggl[ -\frac{1}{2 \beta_2}  \Biggl( \beta_1 
+ {\cal B} 
\tan \biggl( \frac{1}{2}  {\cal B} \rho z' \\
\nonumber  
&-&  \tan^{-1} \Bigl( \frac{\beta_1 + 2\beta_2 \ln \left(E_\tau^i /E_0\right)}{{\cal B}} 
\Bigr)
 \biggr)
\Biggr) \Biggr] E_0  \ ,
\end{eqnarray}
with ${\cal B} = \sqrt{\beta_0 \beta_2 - \beta_1^2}$. 
The survival probability for the three parameter fit to $\beta_\tau$ is
\begin{widetext}
\begin{eqnarray}
\label{eq:psurv-3par}
P_{surv}(E_\tau, E_\tau^i)= \exp \Biggl[ \frac{m_\tau \ e^{(\beta _1-{\cal B})/2 \beta _2 } }{c \tau \rho E_0 {\cal B}} 
&\biggl[& 
\left\{ {\rm Ei} \left(-\frac{\beta _1-{\cal B}}{2 \beta _2}-\ln \Bigl(\frac{E_\tau^i}{E_0} \Bigr) \right)
-{\rm Ei} \left(-\frac{\beta _1-{\cal B}}{2 \beta _2}-\ln \Bigl(\frac{E_\tau^i}{E_0} \Bigr) \right) \right\} \\
\nonumber
&-& e^{{\cal B}/\beta_2}
\left\{ {\rm Ei} \left(-\frac{\beta _1+{\cal B}}{2 \beta _2}-\ln \Bigl(\frac{E_\tau^i}{E_0} \Bigr) \right)
-{\rm Ei} \left(-\frac{\beta _1+{\cal B}}{2 \beta _2}-\ln \Bigl(\frac{E_\tau^i}{E_0} \Bigr) \right) \right\}
\biggr] 
\Biggr] \ ,
\end{eqnarray}
\end{widetext}
where the exponential integral $ {\rm Ei} (x)  =  -\int^\infty_x e^{-t}/t \ dt$.

Setting $\beta_2=0$ and $d\sigma_{CC}(E_\nu)/dE_\tau=\sigma_{CC}(E_\nu)\delta(E_\tau-(1-y_0)E_\nu)$, the expression for the kernel for the effective aperture is 
\begin{eqnarray}
K(E_\nu,D,E_\tau) &=& \frac{N_A\sigma_{CC}(E_\nu)}{\beta_0 E_\tau[1+\frac{\beta_1}{\beta_0}\ln\biggl(\frac{E_\tau}{E_0}\biggr)]}\\
\nonumber 
&\times & e^{-D\sigma_{tot}\rho N_A}({\cal F})^\xi \\
\nonumber &\times &
P_{surv}((1-y_0)E_\nu, E_\tau)
\end{eqnarray}
where 
\begin{eqnarray}
{\cal F}(E_\nu,E_\tau)&\equiv& \frac{\beta_0+\beta_1\ln((1-y_0)E_\nu/E_0)}{\beta_0+\beta_1\ln(E_\tau/E_0)}\\
\xi &\equiv& \frac{\sigma_{CC}(E_\nu) N_A}{\beta_1}\ .
\end{eqnarray}

\section{Weak interactions}

It is convenient to write the differential neutrino-nucleon cross section as
\begin{eqnarray}
\nonumber
d\sigma&=&\frac{G_F^2 s}{4\pi}\Biggl(\frac{M_V^2}{Q^2+M_V^2}\Biggr)^2
\Biggl[ Y_+ F_2^V - y^2 F_L^V\pm Y_- xF_3^V\Biggr]\\
\nonumber
&=& \frac{G_F^2 s}{4\pi}\Biggl(\frac{M_V^2}{Q^2+M_V^2}\Biggr)^2
\Biggl[ 2(1-y)F_S ^V+ F_T ^VY_+ 
\\
&+ &(1-(1-y)^2 )x F_3^V\Biggr]\ ,
\end{eqnarray}
where $F_T^V=(F_L^V+F_R^V)/2$,  and  $2xF_3 ^V= F_L^V-F_R^V$. The relative coupling of the $W$ and $Z$ are absorbed into the definition of
the structure functions for the $Z$.
For weak interactions, $\lambda = L,R,S$ with
\begin{equation}
F_\lambda^V = \frac{Q^2}{4\pi^2} \int_0^1dz\, \int d^2 r|\psi_\lambda^V(z,r)|^2 \sigma_{\rm dip}(x,r)\ .
\end{equation}

For the dipole model \cite{Fiore:2011gx,Barone:1993es,Kutak:2003bd,Arguelles:2015wba},
where $\bar{z} = 1-z$ and $\Qbar = z\bar{z}Q^2 + m_1 ^2 z + m_2^2 \bar{z}$,
\begin{eqnarray}
\nonumber
|\psi_L^V|^2 &=& \frac{N_C}{2\pi^2}\Biggl[ [(g_v^V-g_a^V)^2 \bar{z}^2 + (g_v^V+g_a^V)^2 z^2]\Qbar^2 K_1^2(r\Qbar)\\
\nonumber
       &+& [g_v^V(zm_1+\bar{z}m_2)- g_a^V(-z m_1+\bar{z}m_2)]^2 K_0^2(r\Qbar)\Biggr]\\
\nonumber \\ \nonumber
|\psi_R^V|^2 &=& \frac{N_C}{2\pi^2}\Biggl[ [(g_v^V+g_a^V)^2 \bar{z}^2 + (g_v^V-g_a^V)^2 z^2]\Qbar^2 K_1^2(r\Qbar)\\
\nonumber
       &+& [g_v^V(zm_1+\bar{z}m_2)+ g_a^V(-z m_1+\bar{z}m_2)]^2 K_0^2(r\Qbar)\Biggr]   \\
\nonumber \\ \nonumber
|\psi_S^V|^2 &=& \frac{N_C}{2\pi^2Q^2 }\Biggl[ [ (g_v^V)^2\bigl(2Q^2 z\bar{z} +(m_1-m_2)(zm_1-\bar{z}m_2)\bigr)^2\\
\nonumber
&+&(g_a^V)^2 \bigl( 2Q^2 z\bar{z} +(m_1+m_2)(zm_1+\bar{z}m_2)\bigr)^2]K_0^2(r\Qbar)\\
\nonumber
       &+& [(g_v^V)^2(m_1-m_2)^2+(g_a^V)^2(m_1+m_2)^2] \Qbar^2 K_1^2(r\Qbar)\Biggr]    \\
\end{eqnarray}

\begin{table}[h]
\caption{Couplings for neutrino charged current ($V=W$) and neutral current ($V=Z$) scattering.}
\begin{center}
\begin{tabular}{|c|c|c|c|c|}
\hline
Process & $m_1$ & $m_2$ & $g_v^V$ & $g_a^V$\\
\hline
$V=W$  & $m_d,m_s,m_b$ & $m_u,m_c,m_t$ & 1 & 1\\
\hline
$V=Z $  & $m_d,m_s,m_b$ & $m_d,m_s,m_b$ & $\sqrt{2}(-\frac{1}{2}+\frac{2}{3}\sin^2\theta_W)$ & $-\sqrt{2}/2$\\
\hline
$V=Z $ & $m_u,m_c,m_t$ & $m_u,m_c,m_t$ & $\sqrt{2}(\frac{1}{2}-\frac{4}{3}\sin^2\theta_W)$ & $\sqrt{2}/2$\\
\hline
\end{tabular}
\end{center}
\label{table:dipolegv}
\end{table}

\begin{table}[htp]
\caption{Charged and neutral current weak interaction cross sections with the BDHM parameterization
and Soyez dipole, with slow rescaling $\chi = x(1+(m_i+m_j)^2/Q^2)$, on an isoscalar nucleon target. }
\begin{center}
\begin{tabular}{|c|c|c|c|c|}
\hline
Energy & $\sigma_{\rm CC}^{\rm BDHM}$& $\sigma_{\rm NC}^{\rm BDHM}$  & $\sigma_{\rm CC}^{\rm Soyez}$  & $\sigma_{\rm NC}^{\rm Soyez}$     \\
$[{\rm GeV}]$   & [cm$^2$] & [cm$^2$] & [cm$^2$] & [cm$^2$] \\ \hline
$10^7$ &$ 1.71\times 10^{-33} $& $6.93\times 10^{-34}$ &$ 1.42\times 10^{-33} $& $6.02\times 10^{-34}$\\ \hline
$10^8$ &$ 4.26\times 10^{-33} $& $1.78\times 10^{-33}$ &$ 3.37\times 10^{-33} $& $1.47\times 10^{-33}$\\ \hline
$10^9$ &$ 8.88\times 10^{-33} $& $3.78\times 10^{-33}$ &$ 7.19\times 10^{-33} $& $3.18\times 10^{-33}$\\ \hline
$10^{10}$ &$ 1.63\times 10^{-32} $& $7.03\times 10^{-33}$ &$ 1.41\times 10^{-32} $& $6.35\times 10^{-33}$\\ \hline
$10^{11}$ &$ 2.72\times 10^{-32} $& $1.19\times 10^{-32}$ &$ 2.60\times 10^{-32} $& $1.17\times 10^{-32}$\\ \hline
$10^{12}$ &$ 4.24\times 10^{-32} $& $1.86\times 10^{-32}$ &$ 4.63\times 10^{-32} $& $2.05\times 10^{-32}$\\ \hline
\end{tabular}
\end{center}
\label{table:soyezweak}
\end{table}

\end{appendix}

\newpage 

\bibliography{eloss}

\end{document}